\documentclass[twocolumn,english, reprint, aip]{revtex4-1}
\usepackage[T1]{fontenc}
\usepackage[latin9]{inputenc}
\usepackage{geometry}
\usepackage{xcolor}
\usepackage{pdfcolmk}
\usepackage{babel}
\usepackage{array}
\usepackage{booktabs}
\usepackage{textcomp}
\usepackage{multirow}
\usepackage{amsmath}
\usepackage{graphicx}
\usepackage{SIunits}
\PassOptionsToPackage{normalem}{ulem}
\usepackage{ulem}
\usepackage[unicode=true,
 bookmarks=true,bookmarksnumbered=false,bookmarksopen=false,
 breaklinks=true,pdfborder={0 0 0},pdfborderstyle={},backref=false,colorlinks=false]
 {hyperref}
\hypersetup{pdftitle={Hydration free energies and solvation structures with molecular density functional theory in the hypernetted chain approximation},
 pdfauthor={Luukkonen Sohvi, Maximilien Levesque,  Luc Belloni and Daniel Borgis}}

\makeatletter

\providecolor{lyxadded}{rgb}{0,0,1}
\providecolor{lyxdeleted}{rgb}{1,0,0}

\usepackage[T1]{fontenc}

\newcommand{\be}{\begin{equation}}
\newcommand{\ee}{\end{equation}}
\newcommand{\bea}{\begin{eqnarray}}
\newcommand{\eea}{\end{eqnarray}}
\newcommand{\nn}{\nonumber}

\newcommand{\onehalf}{\frac{1}{2}}

\newcommand{\RR}{\mathbf{r}}
\newcommand{\rr}{\mathbf{r}}

\newcommand{\F}{{\cal F}}
\newcommand{\Rhat}{\hat{{\mathbf r}}}

\newcommand{\PP}{\mathbf{P}}
\newcommand{\EE}{\mathbf{E}}
\newcommand{\DD}{\mathbf{D}}

\newcommand{\Pon}{\Omega(\rr)}

\newcommand{\Ld}{{\cal L}}
\newcommand{\Lm}{{\cal L}^{-1}}

\newcommand{\ite}{{\it i.e., }}
\newcommand{\etal}{{\it et al. }}
\newcommand{\ege}{{\it ,e.g., }}

\newcommand{\Ang}{\textrm{A\kern -1.1ex\raisebox{0.7ex}{$^\circ$}}}

\begin{document}

\title{Dielectric response of confined water films: Insights from classical DFT}

\author{Daniel Borgis}
\email{daniel.borgis@ens.fr}
\affiliation{Maison de la Simulation, USR 3441 CNRS-CEA-Universit\'e Paris-Saclay,
91191 Gif-sur-Yvette, France}
\affiliation{PASTEUR, D\'epartement de chimie, \'Ecole normale sup\'erieure, PSL University,
Sorbonne Universit\'e, CNRS, 75005 Paris, France}
\author{Damien Laage}
\affiliation{PASTEUR, D\'epartement de chimie, \'Ecole normale sup\'erieure, PSL University,
Sorbonne Universit\'e, CNRS, 75005 Paris, France}
\author{Luc Belloni}
\affiliation{LIONS, NIMBE, CEA, CNRS, Universit\'e Paris-Saclay, 91191 Gif-sur-Yvette,
France}
\author{Guillaume Jeanmairet}
\affiliation{Sorbonne Universit\'e, CNRS, Physico-Chimie des \'Electrolytes et Nanosyst\`emes Interfaciaux, PHENIX, F-75005 Paris, France}

\begin{abstract}
We re-examine the problem of the dielectric response of highly polar liquids such as water in confinement between two walls using a simple two-variable density functional theory involving number and polarisation densities.  In the longitudinal polarisation case where a perturbing field is applied perpendicularly to the walls, we show that the notion of local dielectric constant, although ill-defined at a microscopic level, makes sense when a coarse-graining over the typical size of a particle is introduced.  The approach makes it possible to study the effective dielectric response of thin liquid films of various thicknesses in connection to the recent experiments of [Fumagalli \etal, Science, 2018, 360, 1339-1342], and to discuss the notion interfacial dielectric constant.  We argue that the observed properties as function of slab dimension, in particular  the very low dielectric constants of the order of 2-3 measured for thin slabs of $\sim 1 \, nm$ thickness do not highlight any special property of water but can be recovered for a generic polar solvent having similar particle size and the same high dielectric constant. Regarding the transverse polarisation case where the perturbing field is parallel to the walls, the associated effective dielectric constant as a function of the slab dimension reaches bulk-like values  at much shorter widths than in the longitudinal case.
\end{abstract}

\maketitle



\section{Introduction}

The dielectric constant  is a macroscopic concept that relates  the linear response of the polarisation vector to the Maxwell electric field\cite{jackson}.
The derivation of the dielectric constant of bulk fluids from  statistical mechanics principles has a long history starting from the early works of Debye, Onsager and Kirkwood\cite{Onsager1936,Kirkwood1939,Booth1951}, with major advances leading to its modern formulation in the 1970s\cite{Nienhuis1971,ramshaw71,ramshaw77,hansen_theory_2013}. The extension to  inhomogeneous liquids and the necessary conditions to define a local, space-dependent dielectric constant $\epsilon(\rr)$ were given by Nienhuis and Deutch\cite{Nienhuis1971} and re-examined thirty years later  by Ballenegger and Hansen\cite{Ballenegger2003}. Such clear  definition is crucial for the implicit solvent models used \ege in biomolecular simulations to represent the aqueous surrounding medium or for deriving effective electrostatic interaction models based on space-dependent dielectric constants\cite{Roux1999}. That question led to a number of early works trying to characterise  $\epsilon(\rr)$ in the vicinity of biomolecules  or membranes using molecular dynamics (MD) simulations\cite{Simonson1995,Simonson1996,Roux1999}. In 2005, Ballenegger and Hansen presented the first MD simulations of a model polar solvent in  confinement between two repulsive walls  in order to define 
a local  $\epsilon(z)$ rigorously using either linear response or a small perturbing electric field. For a perturbation perpendicular to the walls, they were ledcto conclude that such $\epsilon(z)$ is ill-defined and "is not a useful quantity near the walls". This pioneering work has initiated a number of subsequent MD studies of water in confinement or at interfaces using a realistic atomistic representation of both the solvent and  the confining surfaces\cite{Bonthuis2011,Ghoufi2012,Itoh2015,Schaaf2015,Schlaich2016,Olivieri2021}.
This interest was revived recently by the experimental studies of Fumagalli \etal\cite{Fumagalli2018} who reported local capacitance measurements for water confined between two atomically flat walls separated by various distances down to 1 nanometer. Their experiments were interpreted as revealing "the presence of an interfacial layer with vanishingly small polarisation", that translates into an "anomalously low dielectric constant of confined water".

The dielectric properties of confined water have been the subject of many recent simulation studies, including, {\em e.g.}, refs.~\cite{Zhang2013a,Schlaich2016,Zhang2018e,Motevaselian2020a,Ruiz-Barragan2020,Loche2020a,Motevaselian2020,Mondal2020,Olivieri2021,Qi2021,Ahmadabadi2021,Deissenbeck2023}. However, as already stressed in the early work of ref.~\cite{Ballenegger2005}, the convergence of confined water dielectric properties by MD simulations is very difficult to achieve. Recent developments have been devoted to more efficient methods to compute the dielectric constant\cite{Deissenbeck2023}, or to  analytical theoretical approaches based on a dielectric continuum theory (DCT)\cite{Cox2022} or  a nonlocal field theoretical approach\cite{Monet2021}. Different explanations have been proposed for the observed reduction in the perpendicular dielectric constant of confined water. These include a dielectrically 'dead' interfacial water layer caused by orientational constraints imposed by the interface\cite{Fumagalli2018,Motevaselian2020,Mondal2020}, the disruption of the water hydrogen-bond network at the interface\cite{Ahmadabadi2021}, a dielectric boundary effect\cite{Cox2022}, and an excluded volume effect\cite{Olivieri2021,Noji2022}.

Classical density functional theory (DFT) is a well-founded, efficient  theoretical approach to describe  atomic and molecular fluids at interfaces or in confinement; See \ege Refs~\cite{evans_nature_1979,henderson_fundamentals_1992,dietrich92,biben98,oleksy10,wu_density-functional_2007}. In this article, we re-examine the problem of the dielectric response of highly polar liquids such as water in  confinement between two walls using a two-variable density functional theory, in terms of number and polarisation densities,  that we have  derived and used previously for either a generic dipolar fluid \cite{Ramirez2002,Ramirez2005} or for water\cite{Jeanmairet2013,Jeanmairet2016}. It is a simplified version of the full molecular density functional theory (MDFT) formalism that three of us have been developing for a number of years\cite{Borgis2012,Ding2017,Jeanmairet2019b,Borgis2021}. This simplicity (combined with accuracy as will be seen) makes it possible first  to sort out the important physical variables, secondly  to derive analytical solutions and/or  to provide instantaneous numerical solutions that are exempted from the statistical noise inherent to molecular simulations, and this for as many physical situations as desired. We note that a connected DFT approach was recently  applied to the study of polarisation fluctuations in confined water; the coupling of number and polarisation densities was not considered explicitly, however, with an abrupt number density profile introduced as input. \cite{Noji2022}. Our goal is two-fold: 1) To reproduce at a much simpler level and to re-examine previous MD results concerning the  definition of a local (ill-defined) longitudinal dielectric constant close to a wall or in confinement and to extend this definition to that of  a (well-defined) locally coarse-grained dielectric constant. 2) To contribute to the understanding of the experiments of  Fumagalli \etal  and of the notion of "anomalously low dielectric constant" of water in confinement.

The outline of the paper is as follows. Sec.~\ref{sec:dipolar_functional} introduces our two-variable, number and polarisation density free energy functional. It is applied in Sec.~\ref{sec:1D_confinement} to the microscopic structure and longitudinal dielectric response  of a model  Stockmayer fluid, having  the same bulk dielectric constant as water at similar density, in one-dimensional confinement between two graphene-like surfaces. The response is studied as function of slab thickness from less than a nanometer to micrometers.    Sec.~\ref{sec:water} extends the study to a dipolar representation of SPC/E water and to the transverse response in addition to the longitudinal one. Sec.~\ref{sec:conclusion} concludes.

\section{Free-energy functional for a dipolar liquid}

\label{sec:dipolar_functional}

Before discussing a more complete model of water later on, and  in order to distinguish generic dielectric properties from the specific water properties emerging from its special H-bond structure,  we start  with an ersatz of water, namely a Sockmayer fluid composed of Lennard-Jones (LJ) particles embedding a permanent dipole $\mu$, and  whose density and dielectric constant at ambient temperature are similar to those of water. We take the parameters from the early studies of Pollock and Alder\cite{Pollock1980}: $\sigma_{LJ} = 3.024 \,  \Ang$, $\epsilon_{LJ} = 1.87$ kJ/mol, $\mu = 1.835 \, D$, $\rho=0.0289 \, \Ang^{-3}$ or, in dimensionless units, $T* = k_BT/\epsilon_{LJ} = 1.35, \rho^* = \rho \sigma_{LJ}^3 = 0.8, \mu^* = \sqrt{\mu^2/\epsilon_{LJ}\sigma_{LJ}^3} = 2$. Those parameters yield a dielectric constant $\epsilon = 80$. They also correspond  to a state point considered by Ballenegger and Hansen when studying the dielectric properties of the closely related dipolar-soft-sphere model in confinement\cite{Ballenegger2005}. As shown in Refs~\cite{Ramirez2002,Ramirez2005,Jeanmairet2013,Jeanmairet2016,Levesque2012a}, such dipolar liquid submitted to an external potential can be described accurately by a Helmholtz free-energy functional depending on the local number density $n(\rr)$ and local polarisation density $\PP(\rr)$. This functional can be decomposed into  density and polarisation terms,  $\F = \F_n + \F_P$, with the  density term given by

\begin{eqnarray} \label{eq:Fd}
\F_n[n] &= &k_BT \,\int d\RR \,  \left[ n(\rr) \ln(\frac{n(\rr)}{n_0})-n(\rr)+n_0 \right]  \nn \\
&-& \frac{k_BT}{2} \int d\rr_1 d\rr_2 \, \Delta n(\rr_1) \, c_s(r_{12}) \Delta n(\rr_2)   \nn \\
&+&  \int d\rr\, n(\rr)\, V_{0}(\rr)   + \F_B[n(\rr)]
 \label{eq:Fn}
\end{eqnarray}
where $n_0$ is the fluid bulk number density. $V_0(\rr)$ represents the external LJ potential exerted at point $\rr$. $\F_B[n(\rr)]$ is the so called bridge functional, that we take here as a hard-sphere (HS) bridge functional based on fundamental measure theory\cite{rosenfeld_free-energy_1989,roth_fundamental_2002,roth-review10}, using the Kierlik-Rosinberg scalar version\cite{kierlik_free-energy_1990,Levesque2012} and a reference HS diameter defined conventionally as 
$d_{HS} = \sigma_{LJ}(1+0.298T^*)/(1+0.3316T^* + 0.001048T^{*2}$\cite{Fu2015}.
 The polarization part of the functional reads
 \begin{eqnarray}
\F_P[n,\PP] &=&
k_B T \int d\rr \, n(\rr) \times  \nn \\
& & \left( \ln \left[\frac{\Lm(\Pon)}{\sinh(\Lm(\Pon)} \right] + \Pon \,\Lm(\Pon) \right) \nn \\
&- &   \int d\rr \, \PP(\rr) \cdot {\mathbf E}_0(\rr)  -   \int d\rr_1  \, \PP(\rr_1) \cdot  \EE_{exc}(\rr_1) 
 \label{eq:Fpol}
\end{eqnarray}
with $\Omega(\rr) = P(\rr)/\mu n(\rr)$ and $P(\rr) = |\PP(\rr)|$. The first term represents the ideal free energy for an ensemble of non interacting dipoles submitted to an external electric field; there  $\Ld$ designates the Langevin function and $\Lm$ its inverse. $\EE_0(\rr)$ is the external electric field at point $\rr$. The excess electric field $\EE_{exc}(\rr_1)$ is defined by 
\bea
\EE_{exc}(\rr_1)&= &   \onehalf \frac{k_BT}{\mu^2} \int d\rr_2 \, [  c_\Delta(r_{12}) \PP(\rr_2)   \nn \\
&+&     \, c_D(r_{12}) \, \left( 3 \Rhat_{12} \, (\mathbf{P}(\rr_2)\cdot\Rhat_{12})-  \mathbf{P}(\rr_2) \right)  ] 
\label{eq:Eexc}
\eea
where $\hat{\rr}_{12} = \rr_{12}/r_{12}$. In eqs~\ref{eq:Fn} and \ref{eq:Eexc}, $c_S(r), c_\Delta(r), c_D(r)$ represent the spherical and dipolar spherical-invariant projections of the angular-dependent direct correlation function of the bulk liquid at density $n_0$. Those functions are inputs in the theory and are obtained  from  a preliminary simulation of the bulk fluid. See Ref.~\cite{Ramirez2002,Jeanmairet2016} for their behaviour in direct  and Fourier space.

The equilibrium number density and polarisation density are obtained by minimisation of the functional with respect to both $n(\rr)$ and $\PP(\rr)$. Minimisation of $\F_P$ with respect to  $\PP(\rr)$ for a given $n(\rr)$ gives
\be
P(\rr) = \mu n(\rr) \Ld(\beta \mu |\EE_0(\rr) + \EE_{exc}(\rr)|)
\ee
This accounts for dipolar saturation at high local electric fields. It does so at a fully  microscopic level compared to the coarse-grained dipolar Poisson approach of Berthoumieux \etal\cite{Berthoumieux2021}. For small external fields, the ideal free energy in eq.~\ref{eq:Fpol} can be developed at dominant order in polarisation
\be
\F_P^{id}[n,\PP] = \frac{1}{2} \int d\rr \, \frac{\PP(\rr)^2}{\alpha_d n(\rr)}
\label{eq:FPid}
\ee
where $\alpha_d = \mu^2/3 k_BT$ is the  thermal polarizability of a permanent dipole in a field. In that case,  minimisation yields a linear  relation between $\PP(\rr)$ and $\EE_0(\rr)$, with  indeed a nonlocal response function.

\section{Confinement in a one-dimensional slit pore}

\label{sec:1D_confinement}

\begin{figure}[h]
   \centering
        \includegraphics[width=0.4\textwidth]{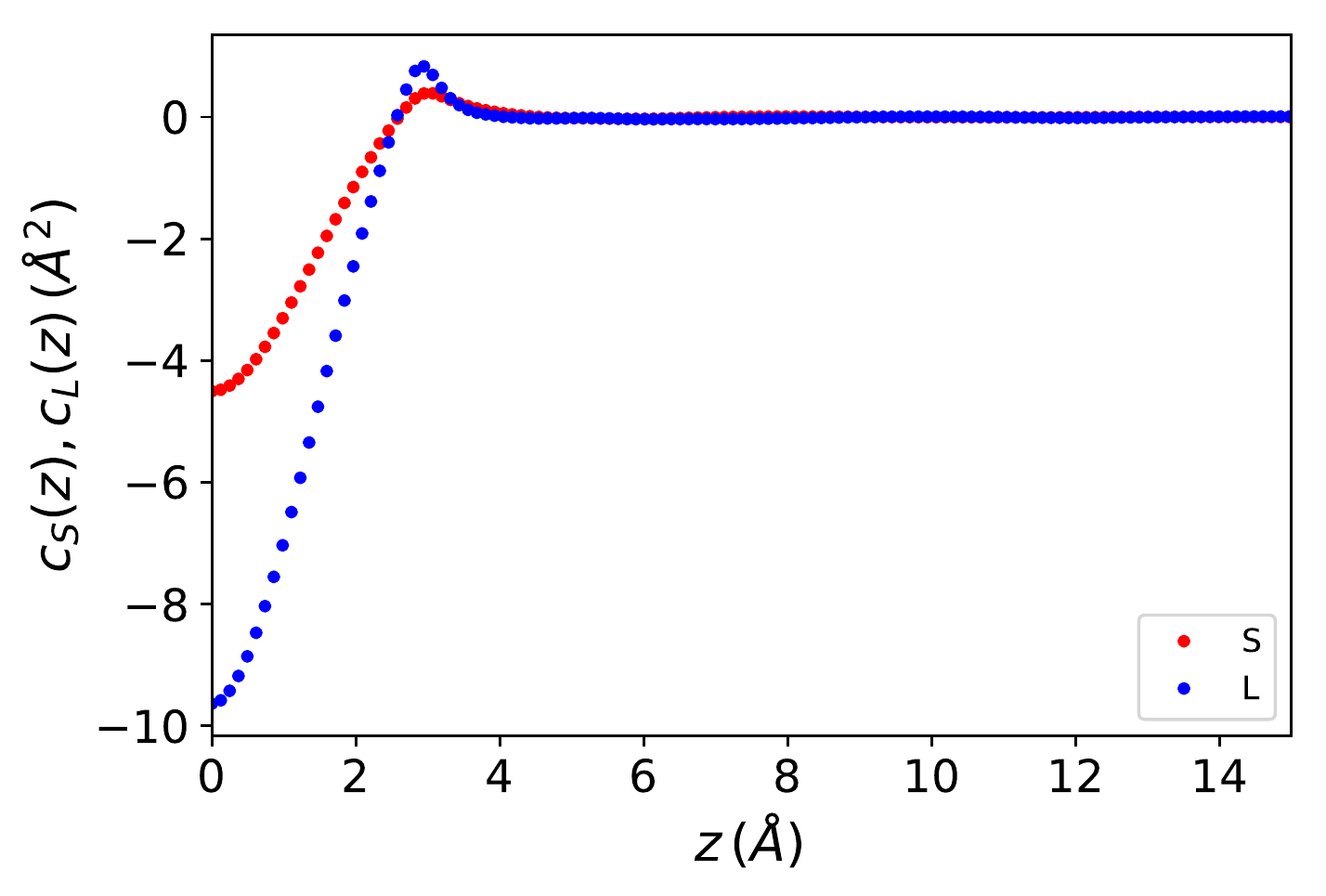} \\
    \caption{One-dimensional direct correlation functions  for the Stockmayer fluid model described in the text, and entering in eqs~\ref{eq:Fn1D}-\ref{eq:FP1D}}
    \label{fig:cS_L_of_z}
\end{figure}

In order to mimic the experimental setup of Ref.~\cite{Fumagalli2018}, as well as to follow the simulation conditions of Ballenegger and Hansen\cite{Ballenegger2005}, we consider a model of one-dimensional slit pore composed of 2 graphene-like plates in the $x$-$y$- plane separated by a distance $h$ along $z$. As in Ref.~\cite{Ballenegger2005}, the external potential $V_0(z)$ exerted by the two walls results from the x-y integration of
a 3D-Lennard-Jones potential. It is of the 9-3 type, with parameters pertinent to carbon-water interactions
\be
V_0(z) = \frac{4\pi}{3} \epsilon_w \left[ \frac{\sigma_w^9}{15z^9} + \frac{\sigma_w^9}{15(h-z)^9}  -\frac{\sigma_w^3}{2z^3} -\frac{\sigma_w^3}{2(h-z)^3} \right]
\ee
with $\sigma_w = 3.9 \, \Ang$ and $\epsilon_w = 2.6 $ kJ/mol.  An external electric field $E_0(z)$ is applied along the perpendicular $z$-direction. For such a 1D-geometry, the polarisation field is so-called longitudinal (\ite aligned with the electric field in  $q$-space), and the two direct correlation functions $c_\Delta(q),  c_D(q)$  (zeroth- and second-order Hankel transform of $c_\Delta(r), c_D(r)$, respectively) reduce in $q$-space to a single longitudinal function 
$c_L(q) =  c_\Delta(q) +2c_D(q)$\cite{Jeanmairet2016}. The two components of the functional $\F = \F_n + \F_P$ of eqs~\ref{eq:Fn}-\ref{eq:Eexc} can be written per surface area in the form
\bea
\F_n[n] &= &k_BT \,\int dz \,  \left[ n(z)  \ln(\frac{n(z)}{n_0})-n(z)+n_0 \right]  \nn \\
&-& \frac{k_BT}{2}
 \int dz_1 dz_2 \, \Delta n(z_1) \, c_{S}(z_{12}) \Delta n(z_2) \nn \\
 &+ &  \int dz \, n(z)\,  V_{0}(z) + \F_B[n(z)]   \label{eq:Fn1D} \\
 \F_P[n,P] &=& k_BT\int dz \, n(z) \times \nn \\
 & & \left[ \ln \left(\frac{\Lm(\Omega(z))}{\sinh(\Lm(\Omega(z))} \right)
+ \Omega(z) \,\Lm(\Omega(z)) \right]  \nn\\
& - &  \frac{1}{6 \alpha_d}  \int dz_1 dz_2 \, c_{L}(z_{12}) \,  P(z_1) P_z(z_2) \nn \\
 &-&   \int dz\, P(z) E_0(z)  \label{eq:FP1D}
\eea
$c_S(z), c_L(z)$ are defined here as the inverse 1D Fourier transforms of the 3D functions $c_S(q), c_L(q)$. They are plotted in Fig.~\ref{fig:cS_L_of_z}. It should be noted that both are short range and vanish  beyond $r \simeq 6 \Ang$. This might be surprising for the polarisation-polarisation contribution $c_L(z)$ since dipole-dipole interactions are a-priori long-range. It is a well-known fact, however, that for a longitudinal polarisation field,   the long-range $1/r^3$ part of the dipolar tensor disappears, and the Maxwell field is rigorously defined  by the local relation $\EE(\rr) = \EE_0(\rr) - 4 \pi \PP(\rr)$. In other words the dielectric displacement is equal to the external field, $\DD(\rr) = \EE_0(\rr)$.

For a small perturbing field $E_0(z)$ and when $n(z)$ is provided independently through the minimisation of eq.~\ref{eq:Fn1D} only (which amounts to neglecting the $n$-$P$ coupling appearing in the ideal term of eq.~\ref{eq:FP1D}), the quadratic form of eq.~\ref{eq:FPid} can be used, turning the minimisation in $P(z)$  to a linear algebra problem that can be solved through matrix inversion. In this linear regime, the nonlocal response can be written in terms of the susceptibility $\chi_0(z_1,z_2)$
\be
P(z_1) =  \int dz_2 \, \chi_0(z_1,z_2) \ \alpha_d  E_0(z_2)
\label{eq:linear_response}
\ee 
with
\be
\chi_0(z_1,z_2) =  n(z_1) \delta(z_{12}) + \frac{h_L(z_1,z_2)}{3} n(z_1)n(z_2) 
\label{eq:h_L}
\ee
where the longitudinal, inhomogeneous pair distribution function $h_L(z_1,z_2)$ relates to the bulk direct correlation function $c_L(z_{12})$ through an inhomogeneous Ornstein-Zernike (OZ)  relation. See the Appendix for details. 
For a constant electric field $E_0(z) \equiv E_0$ a local response function can  be defined as
\be
 f(z) = 4\pi P(z)/E_0 = 1 - \frac{1}{\epsilon_\perp(z)} 
 \label{eq:fz}
 \ee
 where $\epsilon_\perp(z)$ stands for a local {\em longitudinal} dielectric constant and formally
\be
f(z_1) = 4\pi  \alpha_d n(z_1) \left( 1 + \int dz_2  \frac{h_L(z_1,z_2)}{3} n(z_2) \right)
\label{eq:fz_OZ}
\ee
Variants of  this formula can be readily found in the literature\cite{Ballenegger2003}. A few remarks are worth stating: 1) The pair distribution $h_L(z_1,z_2)$ that enters here is not the bulk one; it depends on both $z_1$ and $z_2$, not  on $z_{12}$  only. The fact that the presence of boundaries modifies the fluid response function with respect to the bulk and makes it depend on the two bodies positions rather than only on their relative distance is familiar to inhomogeneous OZ approaches. This fact was also brought up by David Chandler using a Gaussian field theory of fluids with excluded volumes\cite{Chandler1993}, and his findings were further interpreted in a classical DFT framework\cite{Sergiievskyi2017}. Using the bulk $h_L(z_{12})$ may turn out to be a reasonable approximation, especially with a smooth, coarse-grained $n(z)$ as input as done in Ref.~\cite{Monet2021}.  2) The inhomogeneous fluid density $n(z)$ enters in eq.~\ref{eq:fz_OZ} at two places; the first one indicates that the local response function should be zero where there is no particle, $n(z) = 0$. The second one excludes the nonlocal  contribution to the polarisation response coming  from region where the density is zero, $n(z_2) = 0$. This nonlocal cut-off effect  on the polarisation response near the boundaries was pointed out recently by Olivieri \etal \cite{Olivieri2021}. It is contained in the field theoretical approach of Monet \etal\cite{Monet2021}. 3) Since $h_L(z_1,z_2)$ like $c_L(z_{12})$ is short ranged (see Fig.~\ref{fig:cS_L_of_z}),  the influence of the walls is expected to be short-ranged too, and  the bulk value of $f(z)$ and $\epsilon_\perp(z)$ should be reached after only a few particle diameters from the walls. 
 
From now on, we depart from this linear algebra formulation. The  results presented next were obtained numerically by the joint minimisation of the functional with respect to $n(z)$ and  $P(z)$ in the presence of a small and constant external  field $E_0 = 0.1$ V/nm. We have written a  simple, dedicated Python code for that purpose. For a discretisation of the fields over typically $N= 1024$ points, the minimisation procedure is instantaneous on a laptop (less than a second).

 \begin{figure}[h]
  \centering
        \includegraphics[width=0.35\textwidth]{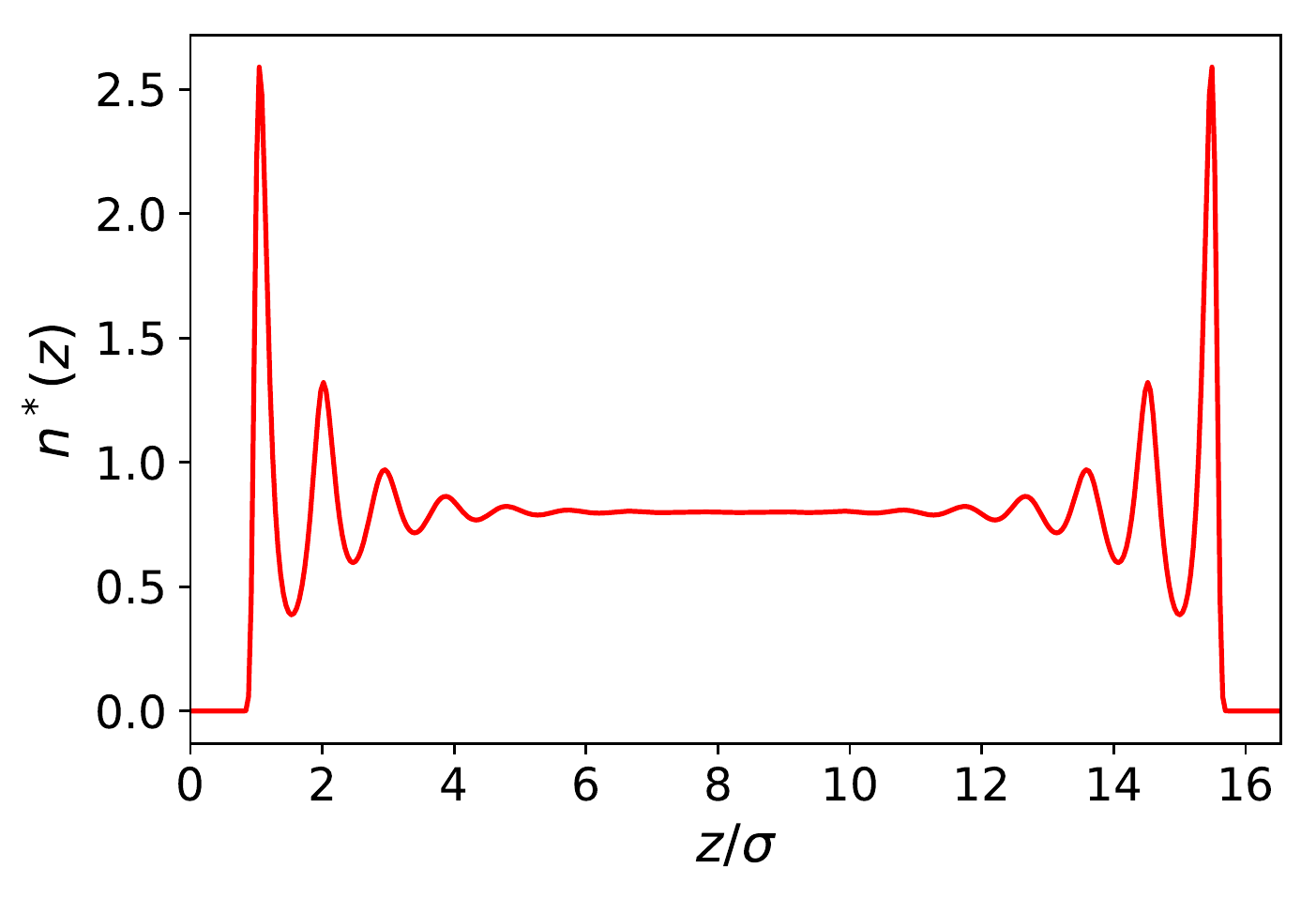} \\
         \includegraphics[width=0.35\textwidth]{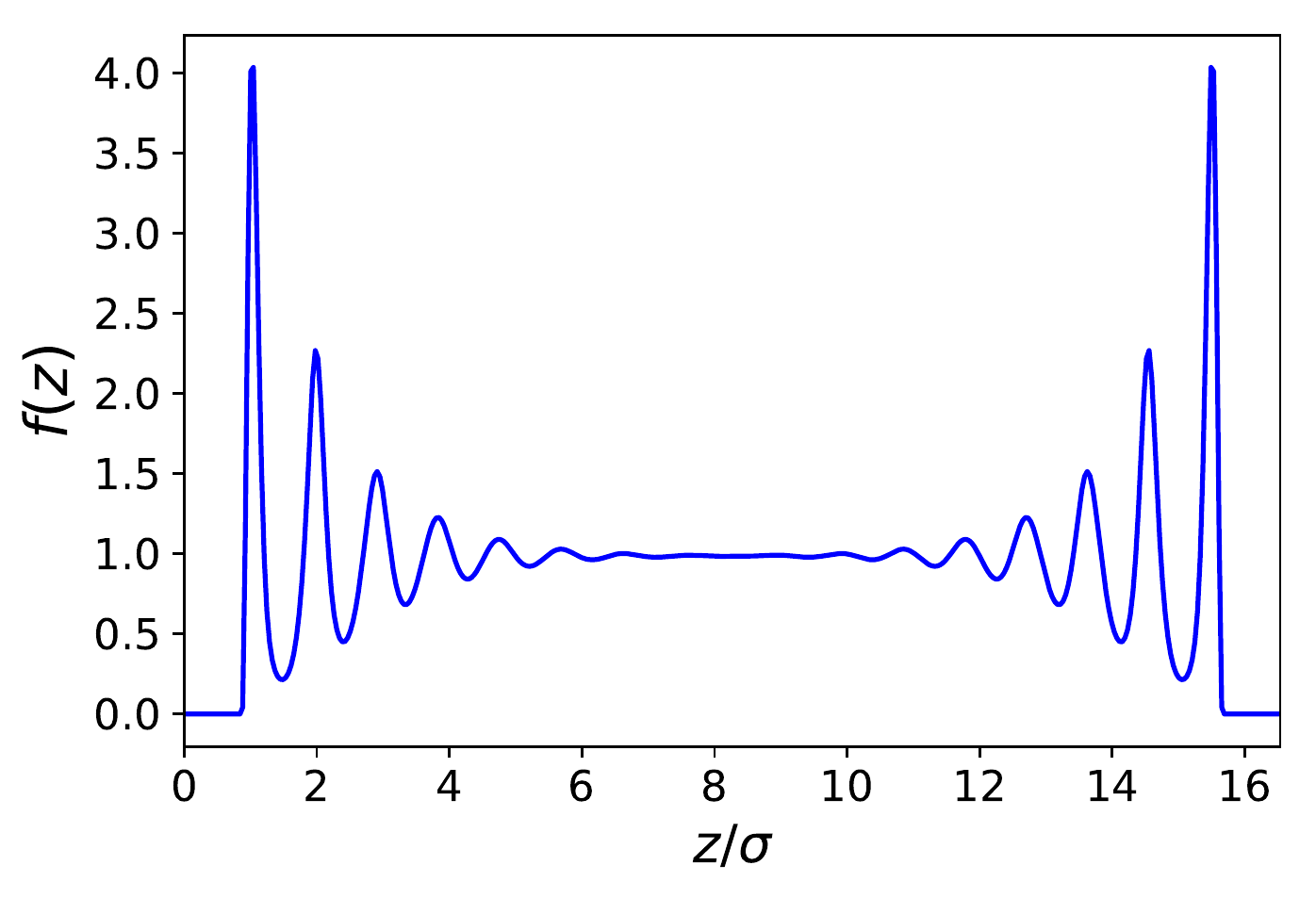} \\
    \caption{Top: Reduced density $n^*(z) = n(z)\sigma_{LJ}^3$ for a slab of width $h=50 \, \Ang$   ($16.5\sigma_{LJ}$). Bottom: Local response function $f(z)=4\pi P(z)/E_0$.}
    \label{fig:nstar_f_h=50}
\end{figure}

\begin{figure}[h]
  \centering
        \includegraphics[width=0.35\textwidth]{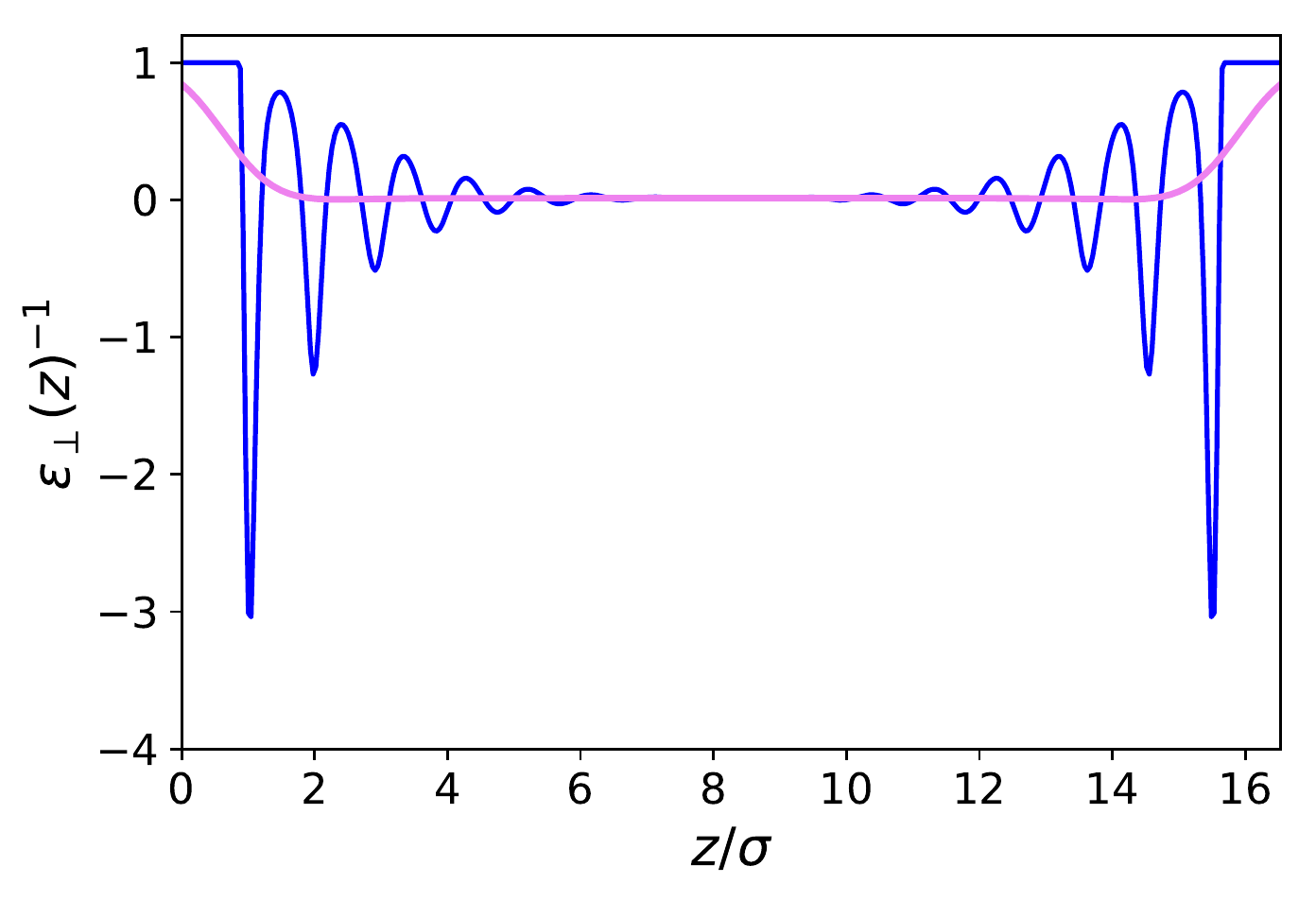} \\
        \includegraphics[width=0.35\textwidth]{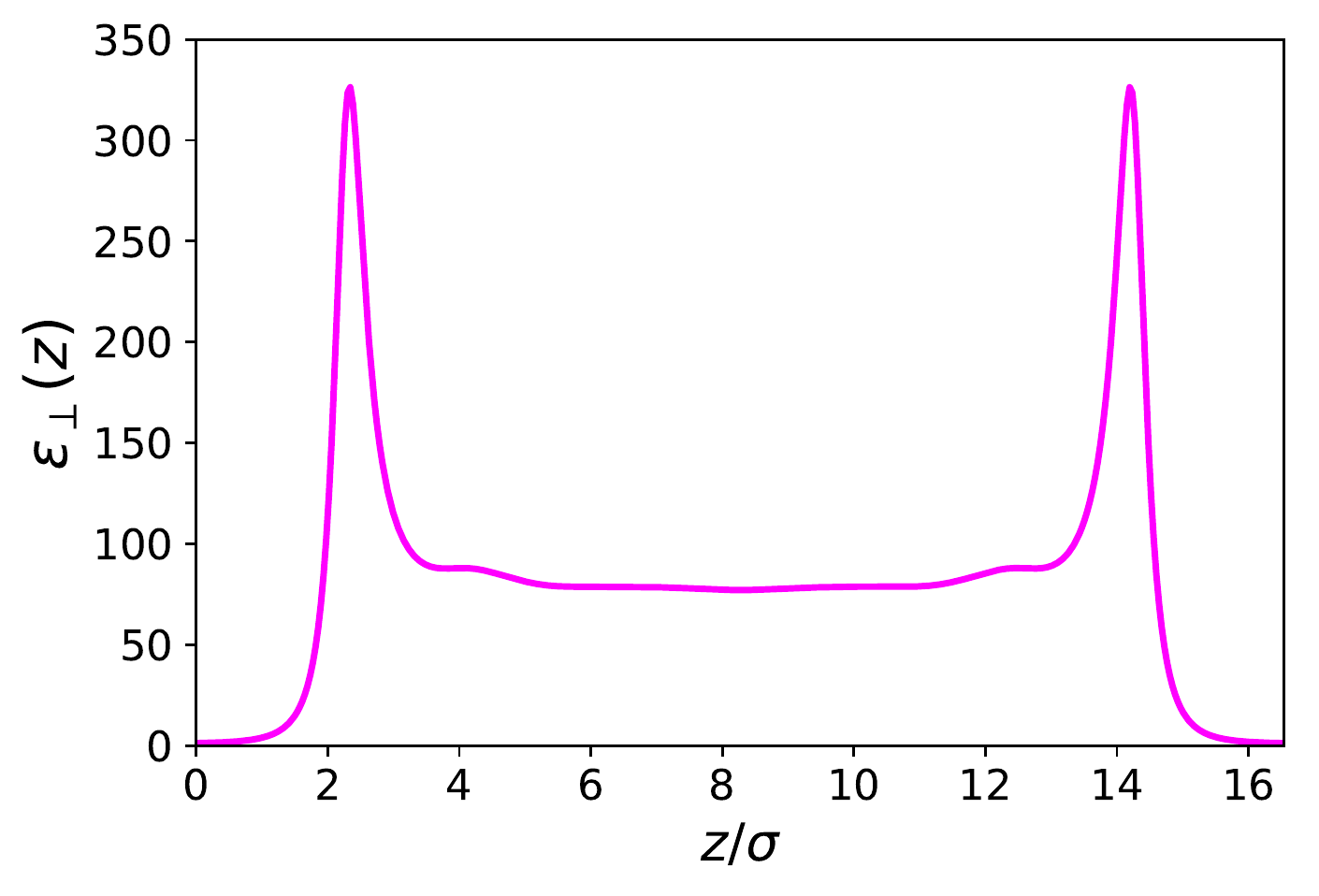}
    \caption{Top: Local inverse dielectric constant $1/\epsilon_\perp(z)$ in a slab of width $h=50 \, \Ang$ (blue curve) and its coarse-grained version $1/\tilde{\epsilon}_\perp(z)$ obtained through eq.~\ref{eq:Pbar} with a coarse-graining length  $\sigma_P = 0.7\sigma_{LJ}$ (violet curve). Bottom: Coarse-grained dielectric constant $\tilde{\epsilon}_\perp(z)$ whereas the corresponding microscopic $\epsilon_\perp(z)$ is ill-defined.}
    \label{fig:one_over_epsilon_h=50} 
\end{figure}

\begin{figure}[h]
   \centering
        \includegraphics[width=0.35\textwidth]{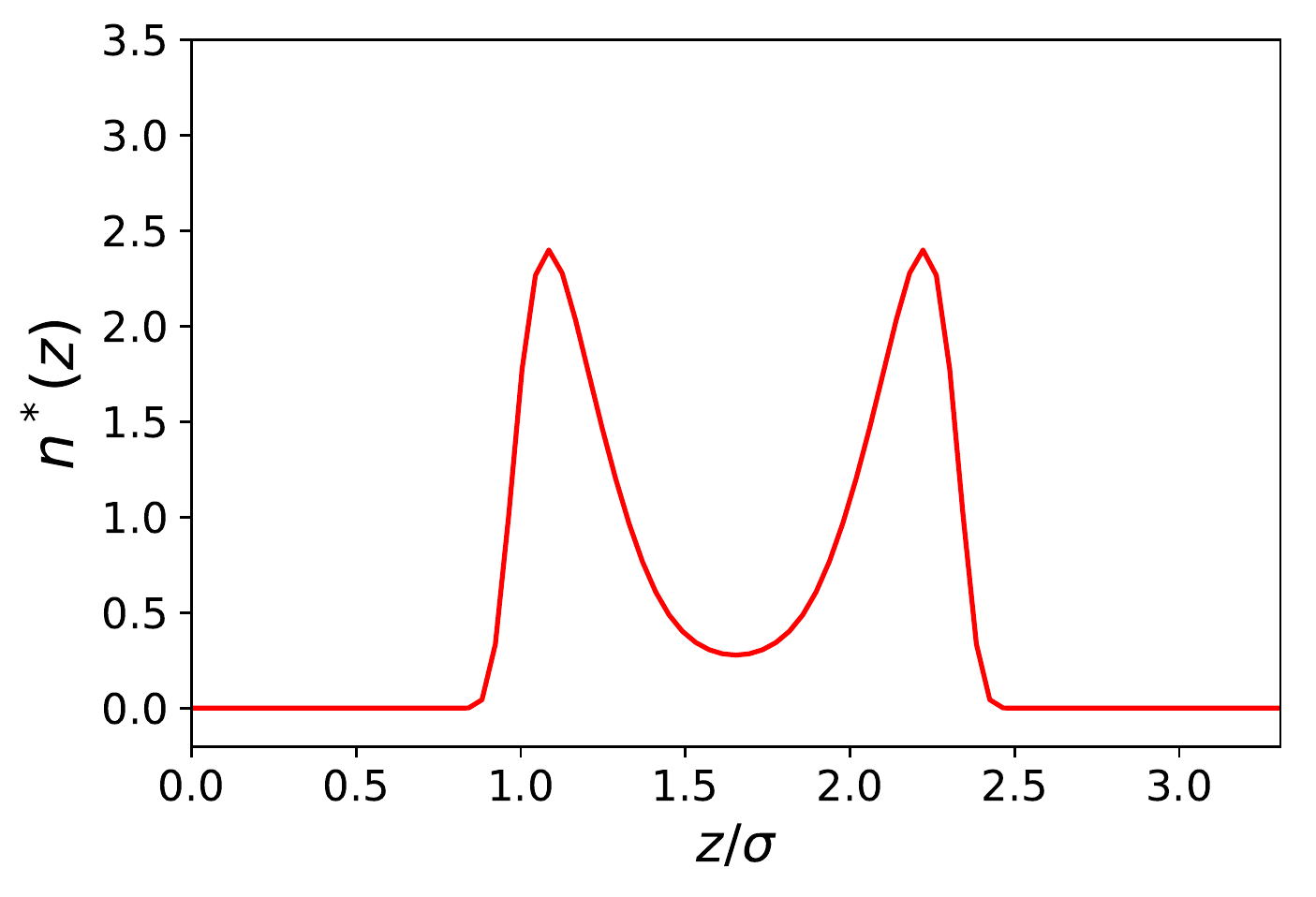} \\
      \includegraphics[width=0.35\textwidth]{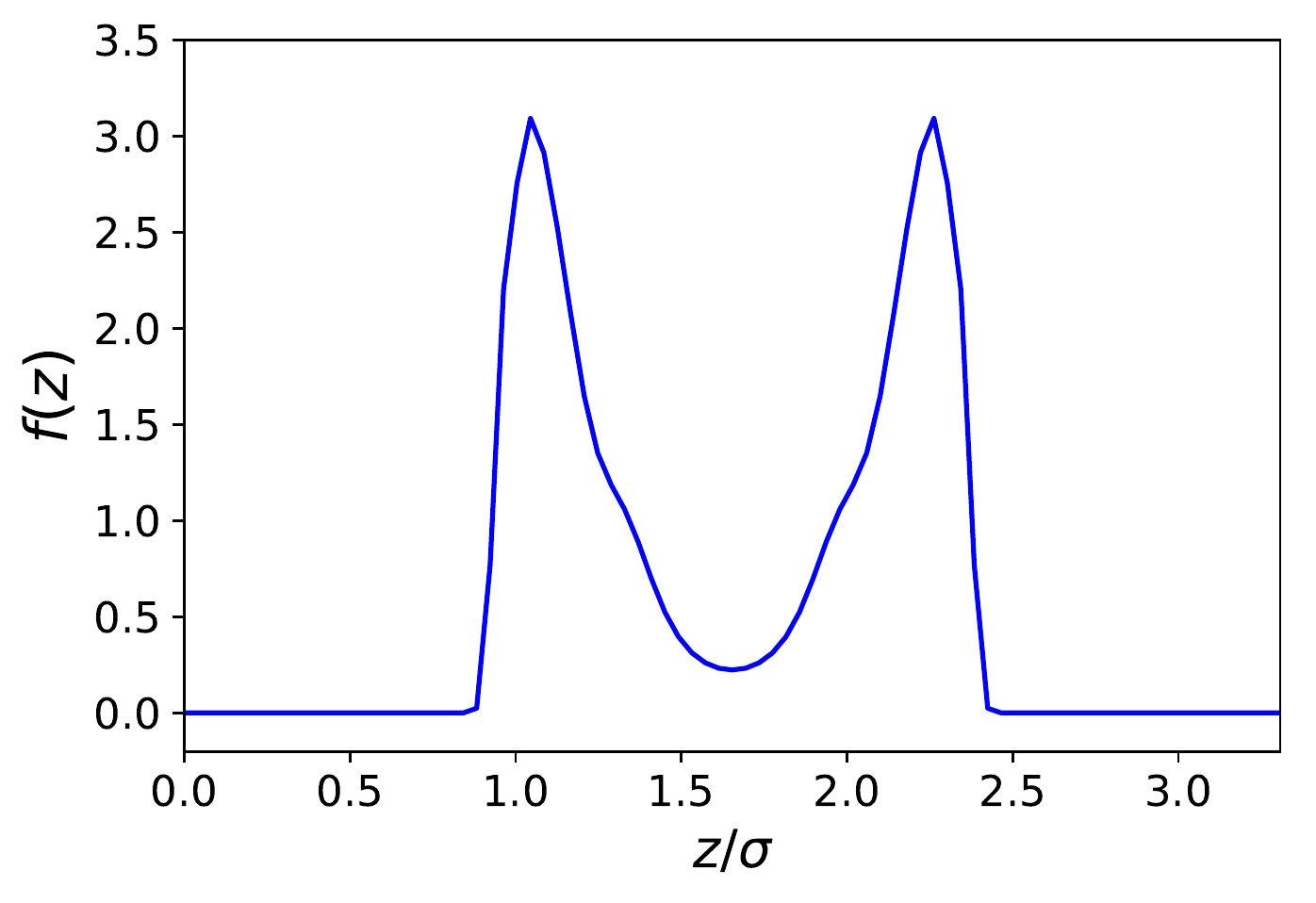}       
    \caption{Same as Fig.~\ref{fig:nstar_f_h=50} for a slab of width $h = 10 \, \Ang$.}
    \label{fig:nstar_h=10}
\end{figure}

\begin{figure}[h]
  \centering
        \includegraphics[width=0.35\textwidth]{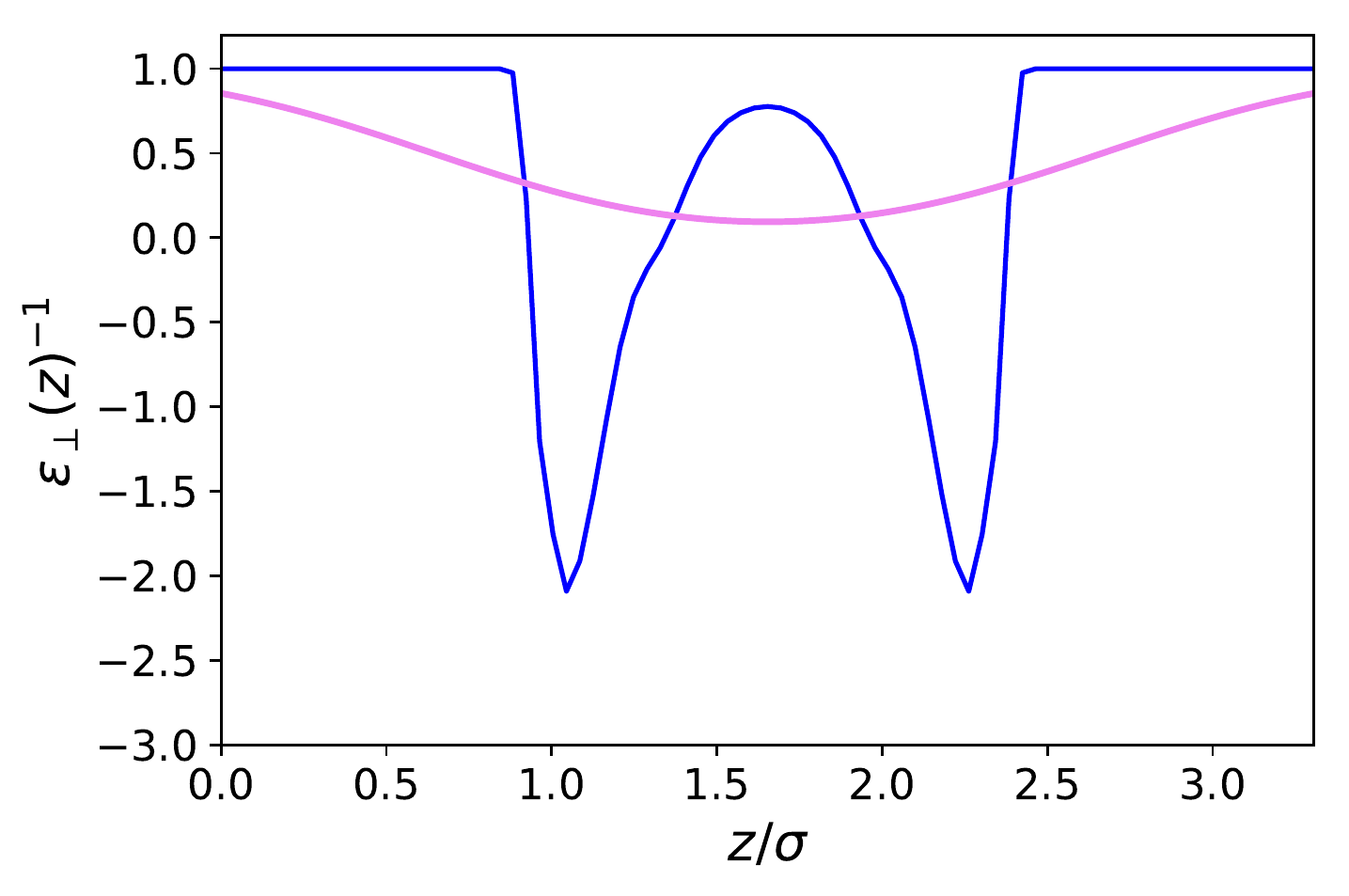} \\
         \includegraphics[width=0.35\textwidth]{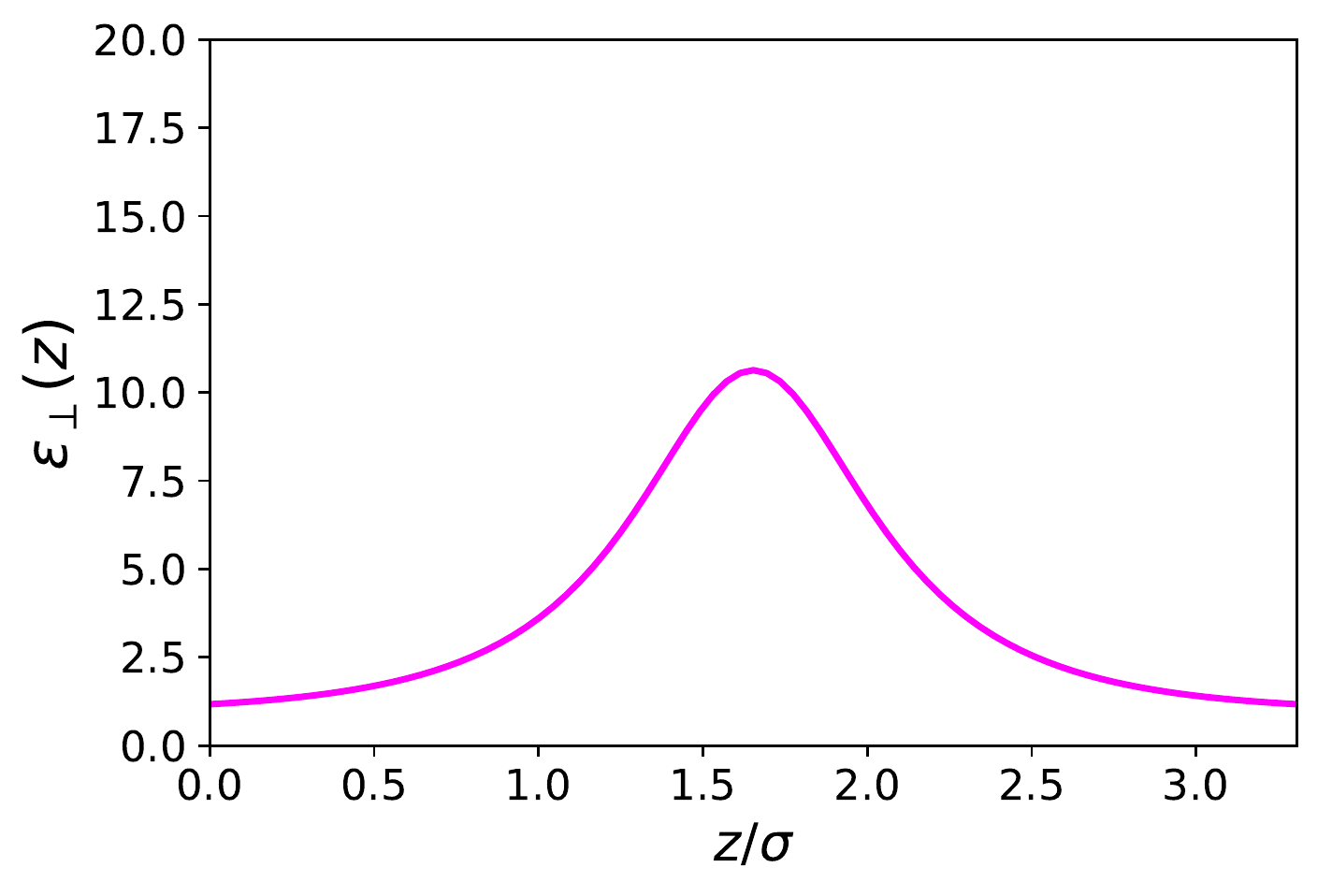}
    \caption{Same as Fig.~\ref{fig:one_over_epsilon_h=50} for a slab of width $h = 10 \, \Ang$. }
    \label{fig:one_over_epsilon_h=10}
\end{figure}

Following the simulations of Ref.~\cite{Ballenegger2005}, we first consider a relatively wide slit of width $h = 50 \, \Ang$ ($16.5 \, \sigma_{LJ}$). We plot the equilibrium density field $n(z)$ as well as the response function $f(z)$ in reduced units in Fig.~\ref{fig:nstar_f_h=50}. Both present strong structural oscillations up to $6$ sigma from the walls. These two curves appear very similar to the ones obtained by Ballenegger and Hansen\cite{Ballenegger2005} via MD-- although their study was mainly focused on the less polar case $\mu^* = 1.2$, the simulations for $\mu^* = 2$ proving very hard to converge. In Fig.~\ref{fig:one_over_epsilon_h=50}-top,   we plot the resulting inverse dielectric constant $1/\epsilon_\perp(z)$ that presents oscillations that span unphysical negative values up to $\sim 6 \, \sigma_{LJ}$ from the walls. That leads Ballenegger and Hansen to conclude that {\em "$\epsilon_\perp(z)$  is not a useful quantity near the walls"}. Here we modulate that judgement by recalling that standard electrostatics is by essence a coarse grained theory, and that one should  rather look at a coarse-grained $\tilde{\epsilon}_\perp(z)$ with a coarse-graining length of at least the size of a particle (this approach was used in ref.~\cite{Ballenegger2005} to smoothen the dipolar fluctuations). Here, this can be formalised by looking at a coarse-grained polarisation field, defined for example by
\be
\tilde{P}(z) = \int dz' w(|z - z'|) P(z')
\label{eq:Pbar}
\ee 
where the weight function $w(z)$ is taken as a normalized Gaussian function with standard deviation  $\sigma_P=\lambda \sigma_{Lj}$, $\lambda$ of order 1. A coarse-grained dielectric constant $\tilde{\epsilon}_\perp(z)$ can be defined from  $\tilde{P}(z)$ exactly as in eq.~\ref{eq:fz}.
The inverse,  coarse-grained,  dielectric constant $1/\tilde{\epsilon}_\perp(\rr)$ corresponding to $\lambda = 0.7$  is plotted  as function of distance in Fig.~\ref{fig:one_over_epsilon_h=50}-top together with the bare microscopic results. This quantity now appears as a smooth curve that does remain strictly positive, so that $\tilde{\epsilon}(\rr)$ itself  is well defined and well behaved; see Fig.~\ref{fig:one_over_epsilon_h=50}-bottom. It presents two peaks at values higher than in the bulk close to the walls; the main feature to be retained, however, is that the bulk value is reached after a few particle diameters ($z \sim 4-5 \, \sigma_{LJ}$) and that there are no long-range effect induced by the walls on the local dielectric constant. In Fig.~\ref{fig:nstar_h=10} and \ref{fig:one_over_epsilon_h=10}, we plot the results corresponding to a much narrower slab with $h = 10 \, \Ang$ 
($\sim 3 \sigma_{LJ}$). It can be seen that only two solvent layers are allowed in-between the plates and that the density $n(z)$ and the polarisation density $P(z)$ remain everywhere far from their bulk values. The coarse-grained dielectric constant $\tilde{\epsilon}_\perp(z)$ displayed in Fig.~\ref{fig:one_over_epsilon_h=10} has a nice and smooth hat shape that reaches a maximum value around $10$ in the middle of the slab, again far below the bulk value.

From the above findings, one can state that the very long range effect, up to a micrometer,  observed for  the measured $\epsilon_\perp$  as function of slab thickness in Ref.~\cite{Fumagalli2018}  cannot be attributed to  any long-range effect of the walls on the local dielectric constant of the liquid. One should look rather at some  effective dielectric  response of the whole slab to the applied potential difference. For our slab model submitted to a constant external field (the so-called dielectric box model of Ref.~\cite{Schlaich2016}) this can be measured  by relating the average polarisation in the slab to the field
\be
\bar{P}(h) = \frac{1}{h} \int_0^h P(z) = \frac{1}{4\pi} \left(1 - \frac{1}{\bar{\epsilon}_\perp(h)}\right) \, E_0
\ee 
which yields according to eq.~\ref{eq:fz}
\bea
\frac{1}{\bar{\epsilon}_\perp(h)} &=& \frac{1}{h} \int_0^h dz \, \frac{1}{\epsilon_\perp(z)} \nn  \\
&\simeq  &  \frac{1}{h} \int_0^h dz \, \frac{1}{\tilde{\epsilon}_\perp(z)}
\label{eq:epsilonbar}
\eea
The second equality holds for the coarse-grained dielectric constant instead of the microsciopic one if the the coarse-graining length is such that $\tilde{P}(0) = \tilde{P}(h) \simeq  0$. Expressing the total electrostatic energy of the device, which includes the self-energy of the electric field between the plates, writing the potential difference between them as
\be
\Delta \Phi(h) = -\int_0^h dz \, E(z) = -\int_0^h dz \, \frac{1}{\epsilon_\perp(z)} \, E_0
\ee
and equating this energy to $1/2 \, C(h) \Delta \phi(h)^2$ yields the effective capacitance
\be
C(h) =  \frac{\bar{\epsilon}_T(h) }{4 \pi h}
\ee
with $\bar{\epsilon}_T(h)$ having the same  definition as in eq.~\ref{eq:epsilonbar}. Measuring the average polarisation in the slab or the effective slab  capacitance are thus equivalent. When the plate-to-plate distance $h$ is large enough as in Fig.~\ref{fig:one_over_epsilon_h=50}, one can divide the device in three regions, two interfacial regions of width
$h_i$ and an intermediate  bulk region of width $h - 2h_i$ where $\epsilon_\perp(z) \simeq \epsilon_{bulk}$. In that case, the choice of $h_i$  results in the definition of an effective dielectric constant $\epsilon_i$ for the interfacial region through
\be
 \frac{1}{\epsilon_i} = \frac{1}{h_i} \int_0^{h_i}dz  \, \frac{1}{\epsilon_\perp(z)} 
 \ee
and the resulting dielectric constant of the whole slab can be written as
\be
 \frac{1}{\bar{\epsilon}_\perp(h)} = 2\frac{h_i}{h}\,  \frac{1}{\epsilon_i} + \left( 1 - 2\frac{h_i}{h} \right) \, \frac{1}{\epsilon_{bulk}}
 \label{eq:one_over_epsilonbar_approximated}
 \ee
 or alternatively
 \be
 \bar{\epsilon}_\perp(h) \simeq \frac{\epsilon_{bulk}}{1 + \frac{2h_i}{h}(\frac{\epsilon_{bulk}}{\epsilon_i}- 1) } 
 \label{eq:epsilonbar_approximated}
 \ee
 Visual inspection of Fig.~\ref{fig:one_over_epsilon_h=50} leads to the choice $h_i \simeq 6 \sigma_{LJ}$ when looking at the bare $1/\epsilon_\perp(z)$, or $h_i \simeq 3 \sigma_{LJ}$
 when looking at the coarse-grained  curve $1/\tilde{\epsilon}_\perp(z)$.
Here we can define $h_i$ unambiguously as the value under which we find that the approximation in eq.~\ref{eq:one_over_epsilonbar_approximated} departs from the exact integral in eq.~\ref{eq:epsilonbar}.  This criterion gives us $h_i \simeq 9 \, \Ang$, and $\epsilon_i \simeq 5$. The approximated formulas \ref{eq:one_over_epsilonbar_approximated}-\ref{eq:epsilonbar_approximated} match completely  the model of 3 capacitors in series that was used  in Ref.~\cite{Fumagalli2018} to interpret the experimental results, except that  here $h_i, \epsilon_i$ are not fitting parameters but follow from a microscopic analysis. We note that the final formula (equation 14) proposed in  the dielectric continuum theory approach of Cox and Geissler\cite{Cox2022}  amounts in eq.~\ref{eq:epsilonbar_approximated} to reduce the interfacial width $h_i$ to the depletion length where the fluid density is zero (roughly $h_i = 2 \, \Ang$ by inspection of Fig.~\ref{fig:one_over_epsilon_h=50}) and to fix accordingly $\epsilon_i = 1$.

\begin{figure}[h]
   \centering
        \includegraphics[width=0.4\textwidth]{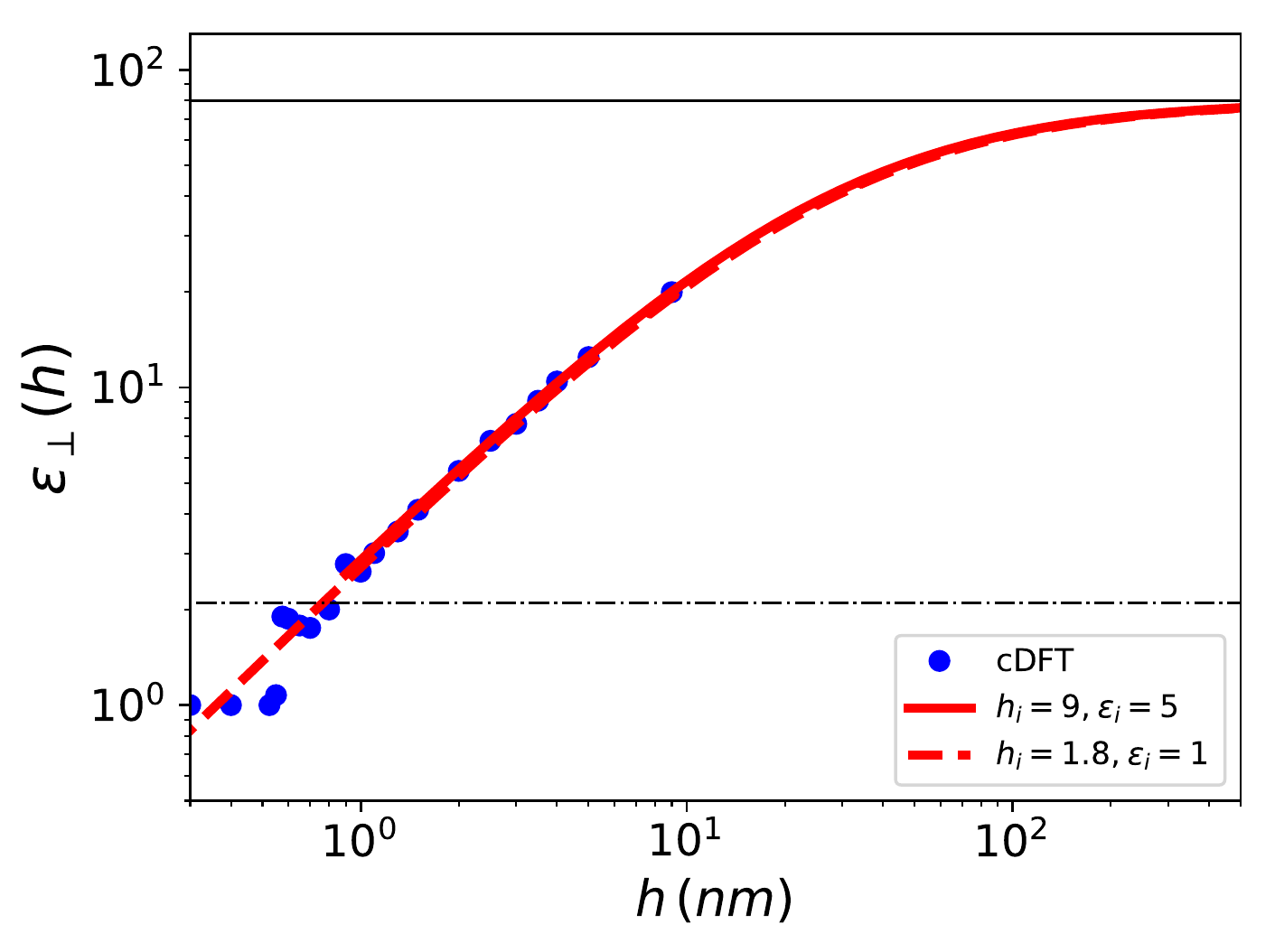} \\
    \caption{Effective dielectric constant computed  using cDFT for the model Stockmayer fluid embedded in a slab of width $h$ as function of $h$. The red solid curve corresponds to the asymptotic formula~\ref{eq:epsilonbar_approximated} with the parameters determined in the text; it starts at $h=h_i$.  The red dashed curve corresponds to the dielectric continuum theory of Cox and Geissler\cite{Cox2022} which amounts in  eq.~\ref{eq:epsilonbar_approximated} to take $\epsilon_i = 1$ while keeping the ratio $h_i/\epsilon_i$ constant. These curves can be compared with the experimental results reported in  Ref.\cite{Fumagalli2018} for water in nano- to micrometric hBN slits.}
    \label{fig:epsilon_perp_vs_h_loglog}
\end{figure}

The  three separated capacitor picture expressed by eq.~\ref{eq:one_over_epsilonbar_approximated}-\ref{eq:epsilonbar_approximated} should not apply when $h < 2h_i$, \ite below  $\sim 20 \, \Ang$. In that case one has to resort merely to numerical integration 
in eq.~\ref{eq:epsilonbar}. For the $h = 10 \, \Ang\,$ case, illustrated in Figs~\ref{fig:nstar_h=10}-\ref{fig:one_over_epsilon_h=10}, the numerical integral yields  a slab-averaged dielectric constant of $\bar{\epsilon}_\perp = 2.3$ ; this is  a surprisingly small value compared to the bulk, that is in line with the experimental findings for water.
In Fig.~\ref{fig:epsilon_perp_vs_h_loglog}, we have plotted $\bar{\epsilon}_\perp(h)$ computed over the range $0-10 \, nm$, together with the asymptotic formula \ref{eq:epsilonbar_approximated} starting from the same microscopic/nanoscopic distances up to the micrometer range. We use the same log-log representation  as in Ref.\cite{Fumagalli2018} for direct comparison. Although our curves correspond to a  a simplified water model  embedded in  a simplified slab (no H-bonds, no electronic polarisation), the similarities with the experimental results are striking. In particular we recover the main feature pointed out by the experimental work: the effective dielectric constant measured in slabs with a thickness in the $1 \, nm$ range is found around 2; this was quoted as an "anomalously low dielectric constant of confined water". Our theoretical work makes it possible to bring some insight to the interpretation of the experimental results.  Indeed the long range behaviour observed for $\bar{\epsilon}_\perp(h)$  has no mystery since, in longitudinal conditions, the measure of the average polarisation  or capacitance yields the integral of $1/\epsilon_\perp(z)$ which, since $1/\epsilon_{bulk} \ll 1$,  gets its main contribution from the boundaries. Very thick slabs are required for the bulk to contribute. This is clear from the slow $h_i/h$ convergence appearing in eq.~\ref{eq:epsilonbar_approximated}.  It should be noted  that this asymptotic formula using the values $h_i, \epsilon_i$ derived above works extremely well even in the $h =1 \, nm$ range, \ite down to separation distances $h < 2 h_i$ where it should not! We can only attribute this to continuity that allows the extrapolation of the curve on a limited range below its validity. Note also that since $\epsilon_{bulk}/\epsilon_i \gg 1$, the results essentially depend on the ratio $h_i/\epsilon_i$ so that, on an empirical ground,  other choices of those parameters are possible to reproduce the average slab dielectric constant. In particular one can take $\epsilon_i = 1$ and $h_i = 9/5 = 1.8 \, \AA$, a value very close  to the one suggested in the dielectric continuum analysis of Cox and Geissler; this is illustrated in Fig.~\ref{fig:epsilon_perp_vs_h_loglog}. Finally, one observes that there is a structuration due to molecular stacking in our results between $h = 0.6$ and  $1 \, nm$.   It is reminiscent, within error bars, of the plateau detected experimentally in the $1 \, nm$ region.

\section{Extension to SPC/E water}

\label{sec:water}
 
\begin{figure}[h] 
   \centering
       \includegraphics[width=0.4\textwidth]{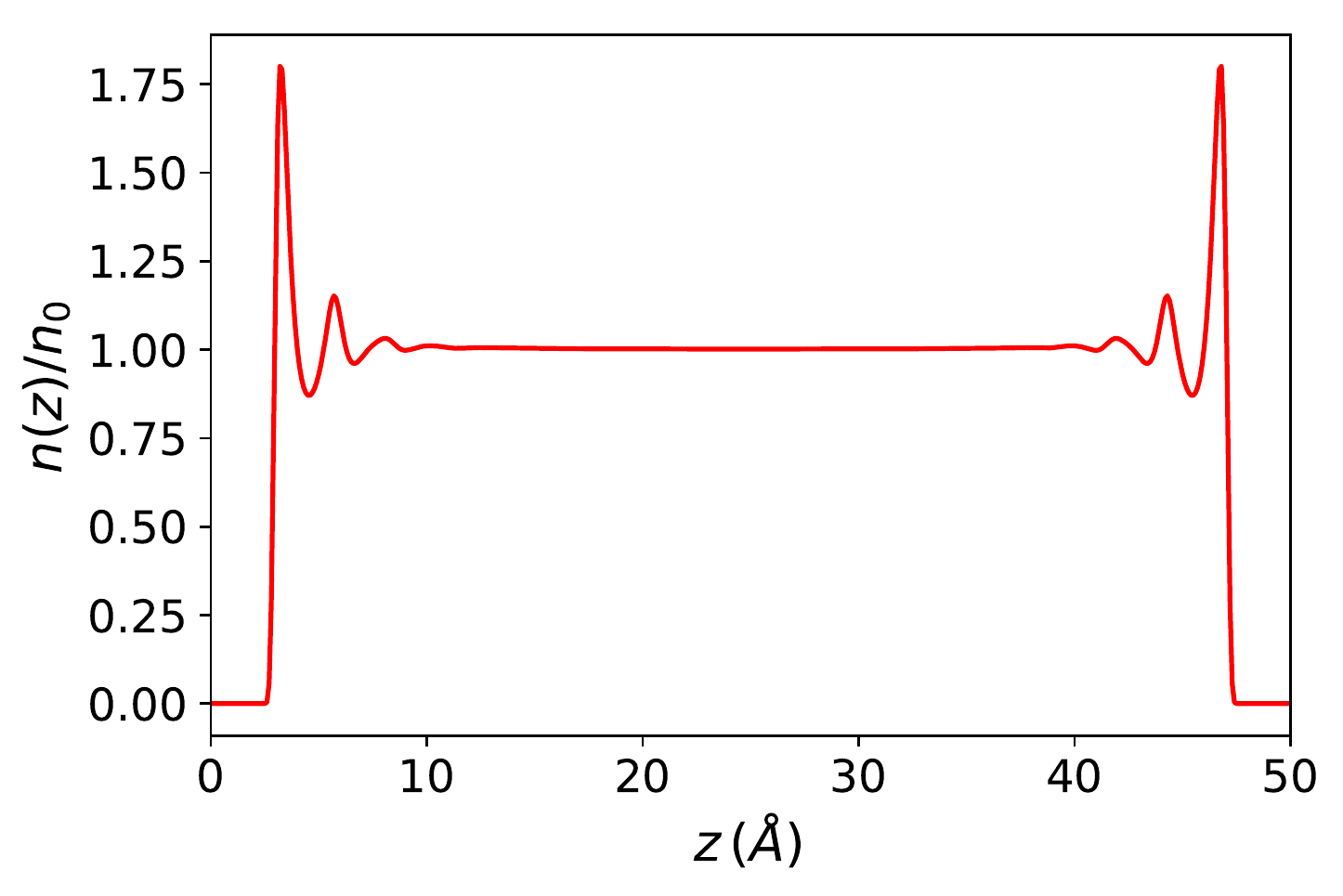} \\
        \includegraphics[width=0.4\textwidth]{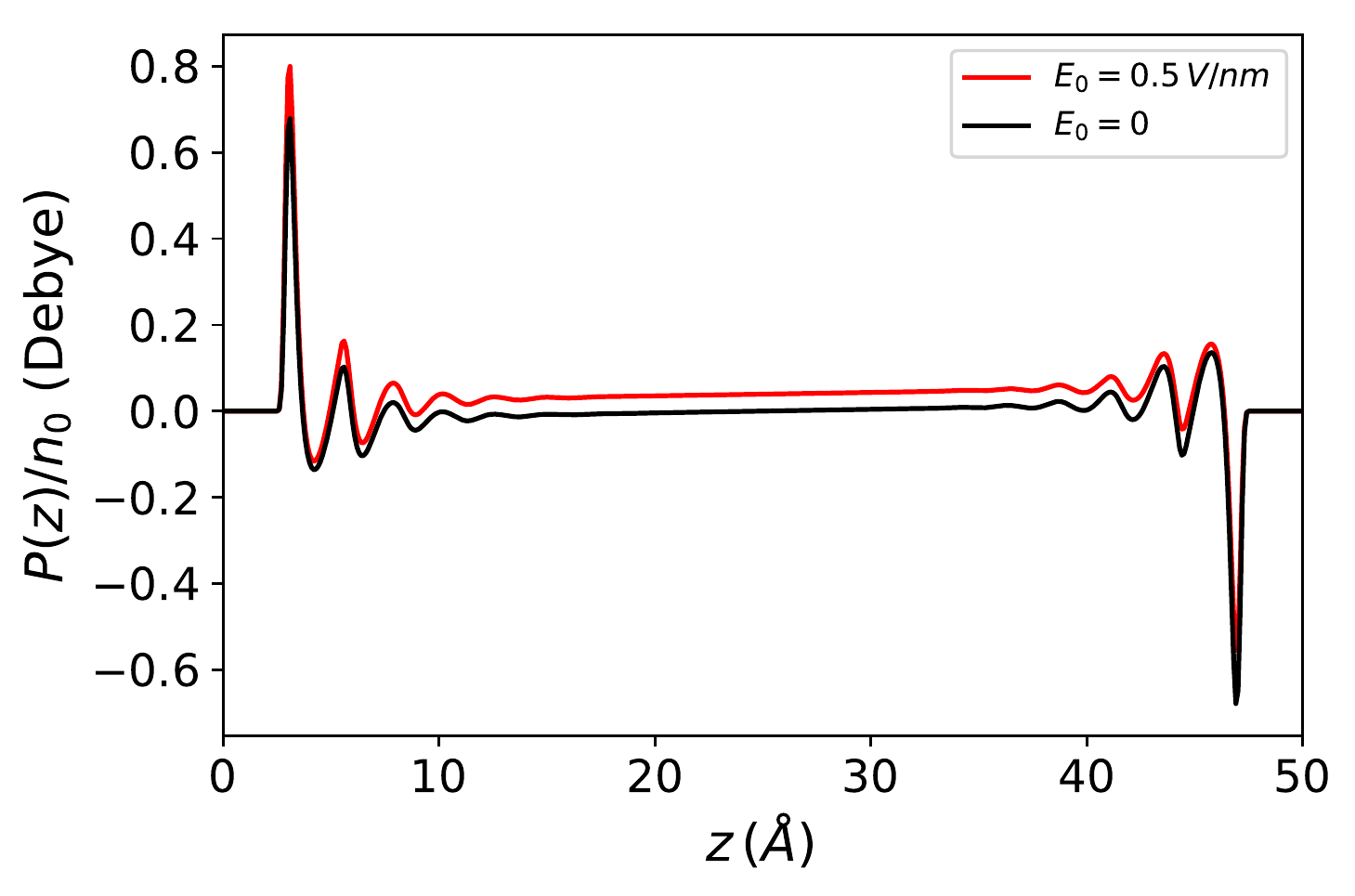} 
    \caption{Top: Reduced density $n^*(z) = n(z)/n_0$ for SPC/E water in a slab of width $h=50 \, \Ang$. Bottom: Polarisation  in the slab with and without an applied external electric field. A spontaneous polarisation exists due to the density/polarisation coupling of eq.~\ref{eq:FnP} which remains the dominant contribution at the interfaces when a field is applied.}
    \label{fig:polarisation_spce}
\end{figure}

 To get even closer to  water, although still at a dipolar level, we extend the previous theory by introducing in the functional the parameters and the $c_S(z), c_L(z)$ direct correlation functions corresponding to SPC/E water. 
 We take the simple weighted density approximation of Ref.~\cite{Borgis2020,Borgis2021} for the bridge functional. Since the symmetry of water is beyond that of a simple dipole,  at least a supplementary density-polarisation coupling has to be introduced  in the functional in the form\cite{Jeanmairet2016}
 \be
\F[n,P] = \F_n[n] + \F_P[n,P] + \F_{nP}[n,P]
\ee
with 
\be
\F_{nP}[n,P]  = - \frac{k_BT}{\mu} \int dz_1 dz_2 \, c_{nL}(z_{12}) \Delta n(z_1) \, P(z_2) 
\label{eq:FnP}
\ee
This  introduces the fact that a spontaneous polarisation exists even  in the slab with zero applied field. The corresponding polarisation profile is anti-symmetrical with respect to the two walls and the integrated  polarisation of the sample is  zero, as it should. This spontaneous polarisation remains prominent when a small to moderate external field is applied. See Fig.~\ref{fig:polarisation_spce} for a large slab with $h=50 \, \Ang$ and also Ref.~\cite{Jeanmairet2019b} where molecular density functional theory calculations were performed  with a molecular representation of the electrodes and a constant voltage applied between the electrodes  rather than an external electric field.

\begin{figure}[h]
   \centering
        \includegraphics[width=0.4\textwidth]{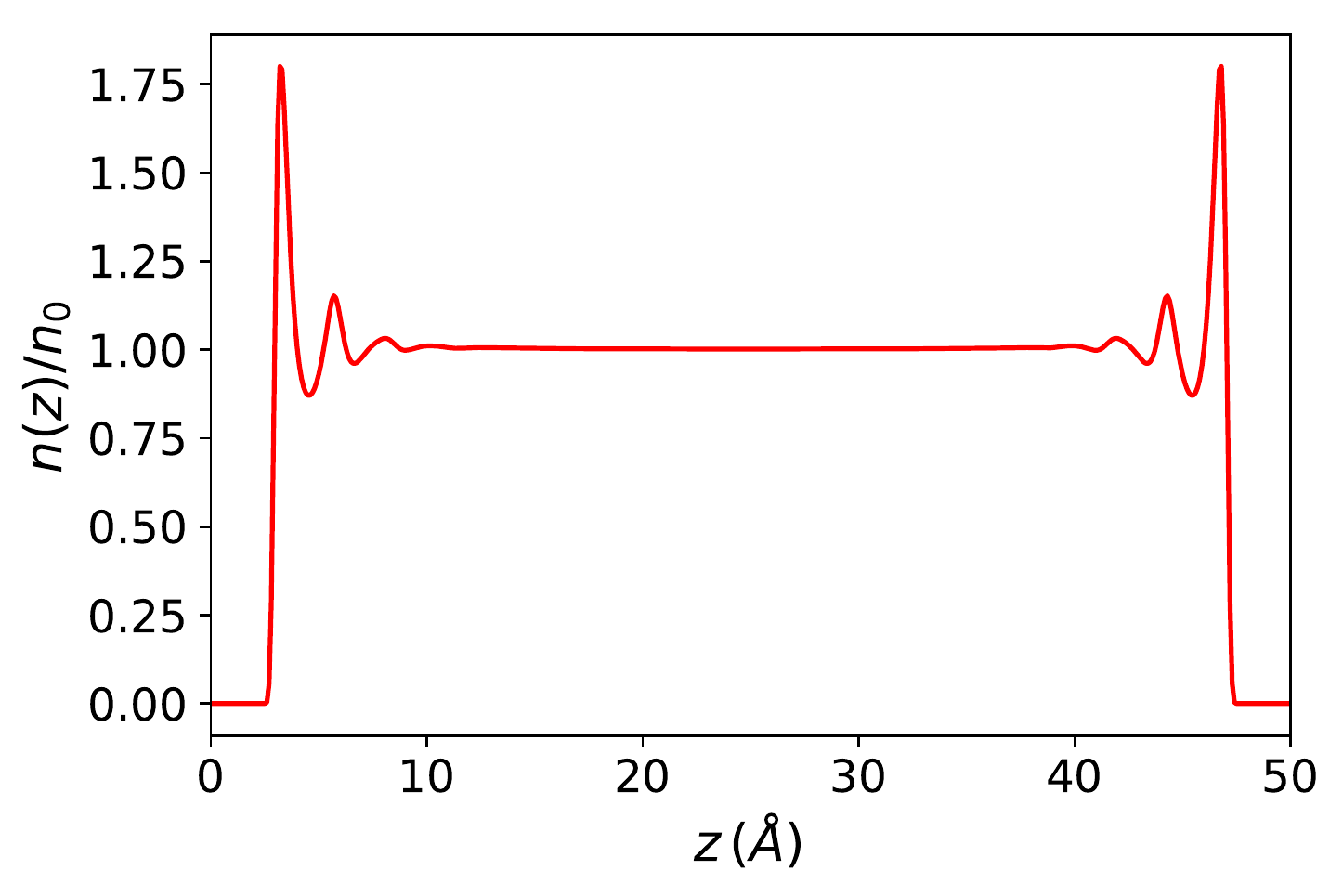} \\
         \includegraphics[width=0.4\textwidth]{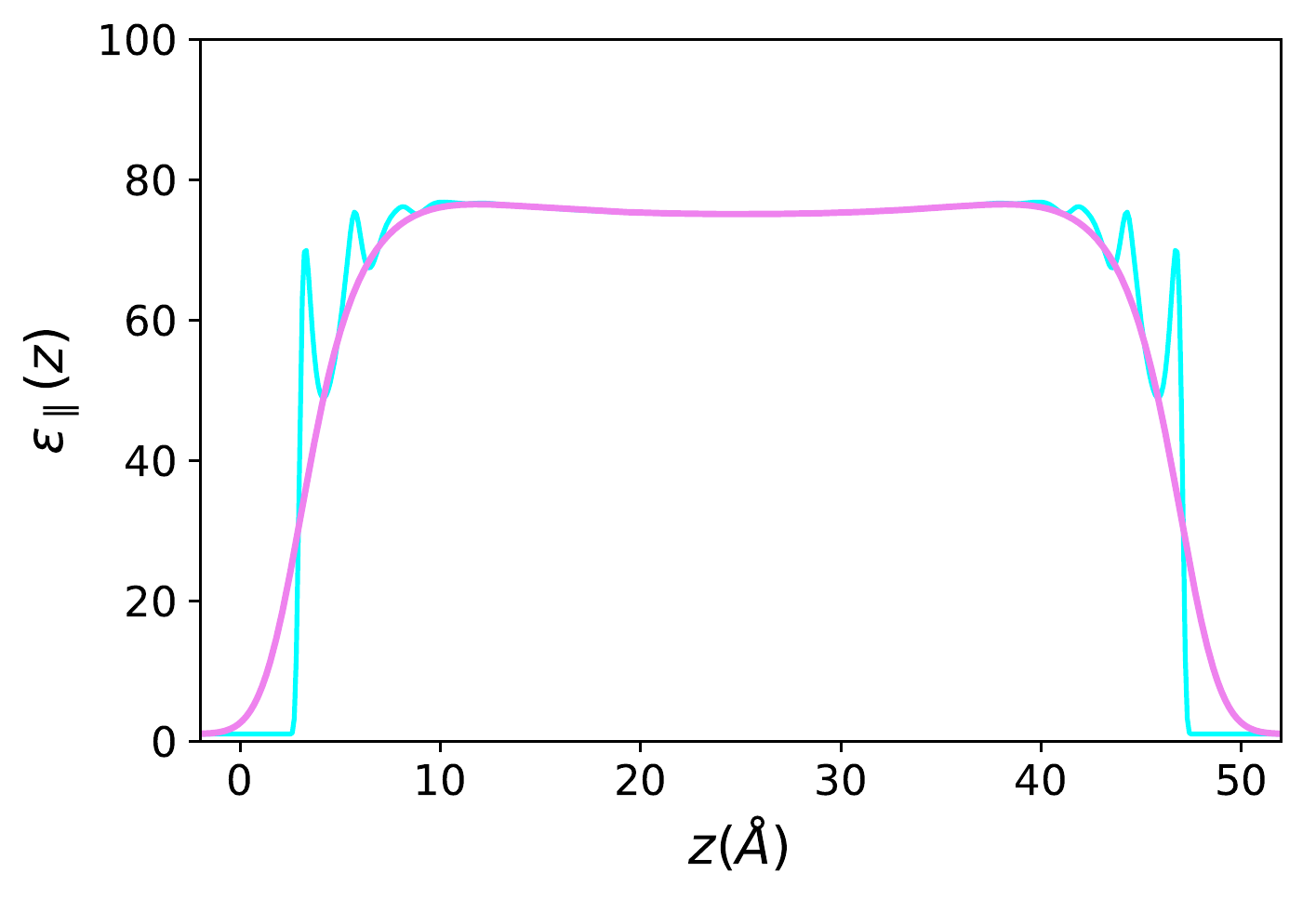}  
    \caption{Top:  (Ill-defined) microscopic perpendicular dielectric constant for SPC/E water in a slab of width $h = 50 \, \Ang$  (in cyan) and its coarse-grained version with a coarse-graining length $\sigma_P=2 \, \Ang$ (in violet). Bottom: Same for the parallel dielectric constant when the field in applied parallel to the plates  ($\sigma_P= 1.5 \, \Ang$).}
   \label{fig:epsilon_perp_h=50_spce}
\end{figure}

On the other hand, the dielectric response $f(z) = 4 \pi (P(z)-P_0(z))/E_0$ appears perfectly symmetrical. In Fig.\ref{fig:epsilon_perp_h=50_spce} , we present this response in terms of the ill-defined, local microscopic constant and of its well-defined coarse-grained version. The latter appears more regular and reaches the bulk more rapidly than in the Stockmayer case of Fig.~\ref{fig:one_over_epsilon_h=50}; this is a sign that the damping of spatial correlations occurs more quickly in water than in a purely dipolar liquid. This is clear also for the number density $n(z)$ when comparing Fig.~\ref{fig:polarisation_spce} to Fig.~\ref{fig:nstar_f_h=50}. The cDFT density in Fig.~\ref{fig:polarisation_spce} appears very similar to the one obtained for a 3D graphene/water/graphene slab of identical width by Olivieri \etal using MD simulations\cite{Olivieri2021} . 

\begin{figure}(h]
   \centering
     \includegraphics[width=0.4\textwidth]{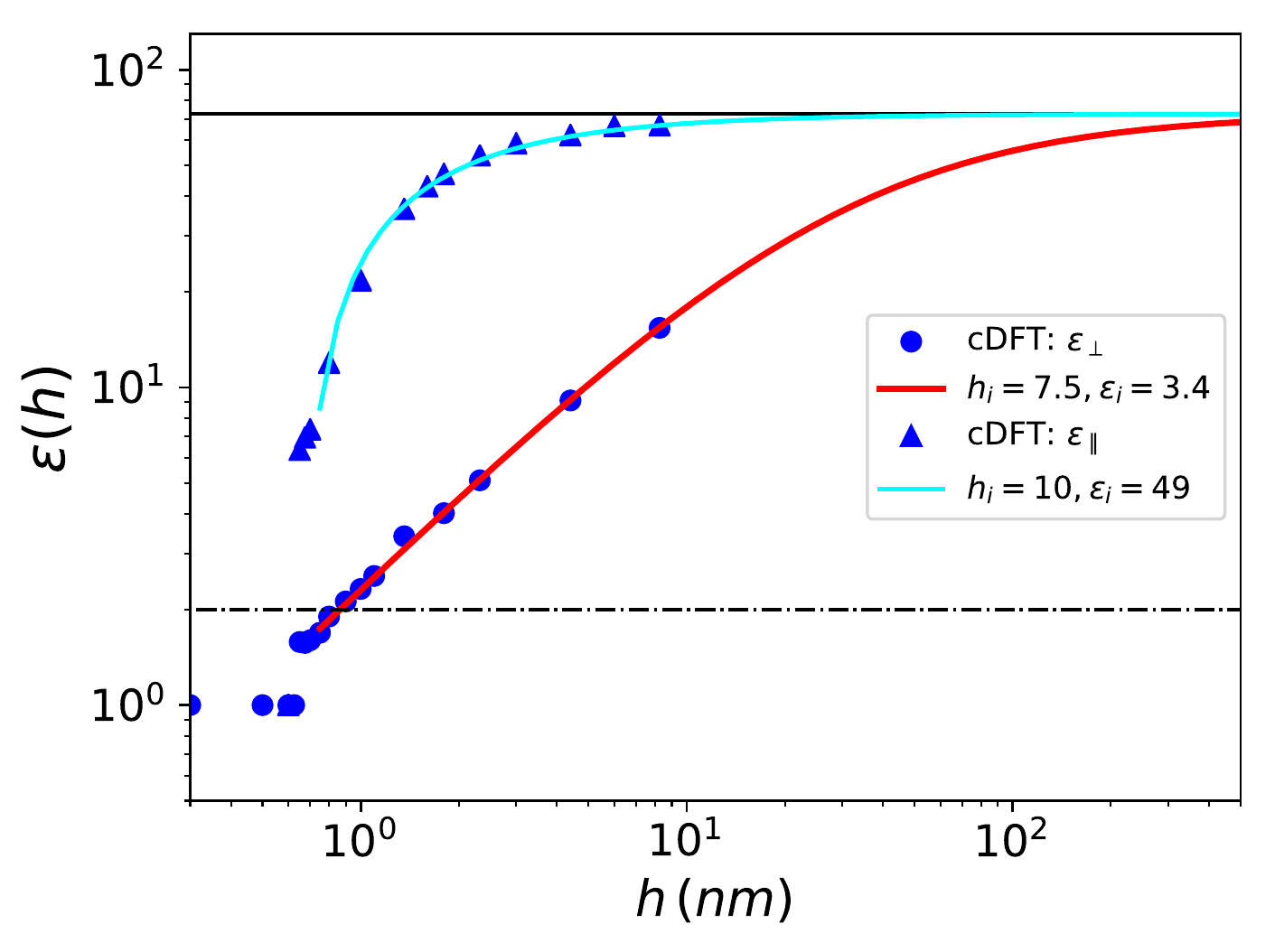}   	 
    \caption{Effective dielectric constant for SPC/E water embedded in a slab of width $h$ as function of $h$. The blue dots are the cDFT results for $\bar{\epsilon}_\perp(h)$  and the red solid curve correspond to the asymptotic formula~\ref{eq:epsilonbar_approximated} with the parameters determined in the text and down to $h=h_i$. The blue triangles and cyan solid curve are the same for $\bar{\epsilon}_\parallel(h)$. }
    \label{fig:epsilon_perp+parallel_vs_h_loglog}
\end{figure}

We present in Fig.~\ref{fig:epsilon_perp+parallel_vs_h_loglog} the curve for $\bar{\epsilon}_\perp(h)$ which is very similar to the one obtained for the Stockmayer solvent, so that identical conclusions can be drawn. No specific property of water emerges, beyond being a polar, molecular fluid of high dielectric constant. This result is consistent with the observation by MD simulations\cite{Motevaselian2020} that confined polar liquids like methanol, acetonitrile and dichloromethane exhibit a dielectric constant reduction similar to that of water. We further note that in our results the dielectric constant reduction of confined water is reproduced by considering exclusively the number and polarisation densities, without requiring orientational constraints imposed by the interface on the water molecules. We find an interfacial width $h_i = 7.5 \, \Ang$ and an associated effective interfacial dielectric constant $\epsilon_i = 3.4$, to be compared to  $h_i = 7.4$ and $\epsilon_i = 2.1$ determined experimentally by Fumagalli \etal\cite{Fumagalli2018}. $h_i$ appears slightly smaller for SPC/E than for the purely dipolar fluid since, as we mentioned,  the spatial correlations in water are shorter range.

Finally, although we are not aware of any experimental results to compare with, we take the opportunity here to study the transverse polarisation case, \ite applying an external electric field in the transverse direction $x$ parallel to the plates. All the DFT formalism developed above remains valid if the longitudinal direct correlation function $c_L(z)$ in eq.~\ref{eq:FP1D} is replaced by the transverse one, $c_T(z)$ defined as the inverse, 1D Fourier transform of $c_\Delta(q) - c_D(q)$\cite{Jeanmairet2016}. The density-polarisation coupling of eq.~\ref{eq:FnP} vanishes  in this case and one resorts to the joint minimisation of the equivalent of the functional in eqs.\ref{eq:Fn1D}-\ref{eq:FP1D} using $c_T(z)$. The response function to a constant field $E_0$ is defined in this case by
\be
f(z) = 4 \pi P(z)/E_0 = \epsilon_\parallel(z) - 1
\ee
The picture is different from that in the longitudinal case since the measure now concerns $\epsilon$ instead of $1/\epsilon$. As seen in Fig.~\ref{fig:epsilon_perp_h=50_spce} for a $50 \, \Ang$-slab, the resulting $ \epsilon_\parallel(z)$ does present oscillations close to the boundaries but remains everywhere positive and  well-defined.
This simple fact was emphasised in the early studies of Ballenegger and Hansen\cite{Ballenegger2005} and confirmed by subsequent MD studies using molecular solvents\cite{Bonthuis2011,Schlaich2016,Ruiz-Barragan2020,Motevaselian2020a,Loche2020a,Olivieri2021}. For a slab of thickness $h$, an effective dielectric constant can be defined as 
\be
\bar{\epsilon}_\parallel(h) = \frac{1}{h} \int_0^h dz \, \epsilon_\parallel(z)
\ee
If the slab is thick enough to distinguish  two interfacial regions from an intermediate  bulk buffer, as illustrated in Fig.~\ref{fig:epsilon_perp_h=50_spce}, the following simple formula pertinent to 3 capacitors in parallel can be inferred by decomposing the integral 
\be
\bar{\epsilon}_\parallel(h) = \epsilon_{bulk} \left[ 1 - 2\frac{h_i}{h}(1-\frac{\epsilon_i}{\epsilon_{bulk}}) \right] 
\label{eq:epsilon_parallel}
\ee
with 
\be
\epsilon_i = \frac{1}{h_i} \int_0^{h_i }dz \, \epsilon_\parallel(z)
\ee
Again $\epsilon_i$ is fixed by the choice of a reasonable $h_i$. Using the same unambiguous selection criterion as before (the minimal $h_i$ that fulfil the asymptotic equation~\ref{eq:epsilon_parallel}), we find $h_i = 10 \, \Ang$ and $\epsilon_i \simeq 50$, \ite a much larger interfacial, effective  value than in the perpendicular case. Besides it can be seen in Fig.~\ref{fig:epsilon_perp+parallel_vs_h_loglog} that the bulk value is reached for slabs of thickness $h \simeq 10 \, nm$, thus much narrower rather than  $\sim 500 \, nm$ necessary for the perpendicular dielectric constant. 

Finally,  let us mention that our DFT results are in overall agreement with previous MD simulations\cite{Bonthuis2011,Schlaich2016,Itoh2015,Motevaselian2020a,Motevaselian2020,Loche2020a,Olivieri2021}. In particular Itoh and Sakuma\cite{Itoh2015} have computed by MD  simulations the perpendicular and parallel effective dielectric constants of three-dimensional graphite/SPCE-water/graphite slabs of various sizes. In spite of our different, simplified modelling of water and  of the surface-water interactions, the DFT results in Fig.~\ref{fig:epsilon_perp_h=50_spce} are in quantitative  agreement with theirs for the few slab geometries that they explored.  

\section{Conclusions}

\label{sec:conclusion}

In this work, we presented a simple two-variable, number/polarisation density functional theory  describing the microscopic structure of polar fluids in confinement, as well as  their microscopic dielectric response to external fields.  For a given slab geometry the numerical solution is obtained  instantaneously with a laptop  and compares very well with previous MD simulations of closely related systems\cite{Ballenegger2003,Olivieri2021}. This approach makes  it possible to  model thin water films of various thicknesses and  to mimic  the experimental setup of Ref~\cite{Fumagalli2018}. Our modelling  is indeed incomplete and neglects physical features such as the precise chemical nature of the interface, its three-dimensional roughness, and the electronic polarisability of both the solid surfaces and the solvent. A main conclusion, however,  is that  finding very low effective longitudinal dielectric constants of the order of 2-3 for water in slabs of nanometer  size through capacitance measurements  should not be a special property of water but is true for any generic polar solvent having a high bulk dielectric constant. Fig.~\ref{fig:epsilon_perp_vs_h_loglog} obtained for the Stockmayer solvent or Fig.~\ref{fig:epsilon_perp+parallel_vs_h_loglog} for a dipolar representation of water present close similarities with  the experimental curve in Ref~\cite{Fumagalli2018}. A similar conclusion is reached in Ref.~\cite{Cox2022} which follows a purely macroscopic,  electrostatics route.  On the other hand the definition  of a local, space-dependent longitudinal dielectric constant is found irrelevant at a microscopic level close to the walls but can be inferred  at a molecular coarse-grained level with a smoothing length of the order of the size of  a solvent particle. This local dielectric constant is shown to reach its  bulk value after a short distance $h_i$ from the slab walls, typically $10\, \Ang$ for water.  This  clearly defines a short-range  interfacial solvent region with an effectively low dielectric constant $\epsilon_i$. Since in the longitudinal polarisation case the response concerns $1/\epsilon$ rather than directly $\epsilon$,  incorporating the intermediate bulk region with a $1/\epsilon_{bulk}$ contribution  turns rigorously to a three-capacitors-in-series model described by  formula \ref{eq:one_over_epsilonbar_approximated}. Since the $1/\epsilon_{bulk}$ is small, the low dielectric constant interfacial regions dominate and it requires large slab thicknesses for the bulk to contribute; this  explains the very slow increase of $\epsilon$ with slab thickness. Our theoretical approach brings additional insight to this simple, phenomenological capacitor model. 1) Although it should not apply to slab width below $2 h_i$, twice the interfacial thickness, it  does hold for shorter distances down to $h_i$. We take this as a continuity effect. 2) In our microscopic analysis, the parameters $h_i$ can be defined unambiguously from the microscopic structure and it fixes also the value of the second parameter $\epsilon_i$. Phenomenologically, since the behaviour in eq.~\ref{eq:one_over_epsilonbar_approximated} depend essentially on the ratio $h_i/\epsilon_i$, other choices of parameter couples are possible including the extreme choice $\epsilon_i=1$ (See Fig.~\ref{fig:epsilon_perp_vs_h_loglog}). 3) We do observe a flattening  of the dielectric response around $\bar{\epsilon}_\perp(h)=2$ for slabs below $1 \, nm$, a saturation effect that is reminiscent of the one detected experimentally. In our case, we can relate this non-monotonic behaviour to the interplay between polarisation response and hard-sphere packing when only one or two layers of solvent particles are allowed in the slab. 

We have also studied the complementary case of the transverse response when the perturbing  field is applied parallel to the walls instead of perpendicular. 
In that case the microscopic dielectric constant $\epsilon_\parallel(z)$ is well-defined although presenting some  structural  oscillations close to the walls; those are  smoothed out by coarse-graining over a particle dimension. A three-capacitors-in-parallel model described by eq.~\ref{eq:epsilon_parallel}  is found to apply for slabs above $1 \, nm$ and  the overall, slab capacitance is found to reach the bulk value for slab thickness on the order of $10 \, nm$, \ite much narrower than in the perpendicular case. The inferred interfacial effective dielectric constant is also much higher.

Finally, let us mention that  water in confinement can be described at a much more refined molecular density functional theory level using the full MDFT formalism and its associated MDFT software that includes not only the dipolar symmetry but  all the higher multipolar symmetries and makes it possible also to represent the surface-water interaction at a fully atomistic, 3D level\cite{Jeanmairet2019b}. We have performed such 3D calculations for the same 1D external potentials as above. They provide results that are very similar to those reported in Fig.~\ref{fig:epsilon_perp_h=50_spce}.

\section{Appendix: Discussion of eqs~\ref{eq:linear_response}-\ref{eq:h_L}}

For small perturbing fields such that the linearisation in eq.~\ref{eq:FPid} applies, the minimisation of the polarisation functional in eq.~\ref{eq:FP1D} for a fixed $n(z)$ yields
\be
\int dz_2 \, \chi_0^{-1}(z_1,z_2) \, P(z_2) = \alpha_d E_0(z_1)
\ee
with
\be 
\chi_0^{-1}(z_1,z_2) =  \frac{1}{n(z_1)} \delta(z_{12}) - \frac{1}{3} c_L(z_{12}) 
\ee
which gives by inversion the linear response formula~\ref{eq:linear_response} relating the polarisation to the external field.
Classically, after decomposition of the susceptibility in a self and distinct contribution as in eq.~\ref{eq:h_L}, 
writing 
\be
\int dz_3 \, \chi_0^{-1}(z_1,z_3) \chi_0(z_3,z_2)  = \delta(z_{12})
\ee 
is equivalent to solving the following inhomogeneous Ornstein-Zernike-like integral equation for $h_L$ knowing $c_L$\cite{hansen_theory_2013}
\be
h_L(z_1,z_2) = c_L(z_{12}) + \frac{1}{3} \int dz_3 \,  c_L(z_{13}) n(z_3)  h_L(z_3,z_2)
\ee
The dependence of the ideal part of $ \chi_0^{-1}$ in the local density $n(z_1)$ makes that both $ \chi_0$  and $h_L$ depend on $z_1$ and $z_2$ rather than just $z_{12}$.

\section*{Acknowledgements}
This work was supported by the Agence Nationale de la Recherche, projet ANR BRIDGE AAP CE29.


\end{document}